\documentclass[a4paper,fleqn]{cas-sc}

\usepackage[authoryear]{natbib}

\begin{document}
\let\WriteBookmarks\relax
\def\floatpagepagefraction{1}
\def\textpagefraction{.001}
\shorttitle{An adaptive framework for the axisymmetric pulsar magnetosphere using physics-informed KANs}
\shortauthors{S. Rigas et~al.}

\title [mode = title]{An adaptive framework for the axisymmetric pulsar magnetosphere using physics-informed Kolmogorov-Arnold networks}

\author[1]{Spyros Rigas}[orcid=0009-0009-2352-8709]
\cormark[1]
\ead{spyrigas@uoa.gr}

\credit{Conceptualization, Methodology, Software, Formal analysis, Data Curation, Writing - Original Draft}

\affiliation[1]{organization={Department of Digital Industry Technologies, School of Science, National and Kapodistrian University of Athens},
                postcode={34400},
                postcodesep={},
                city={Psachna},
                country={Greece}}

\author[2]{Ioannis Contopoulos}[orcid=0000-0001-6890-4143]
\ead{icontop@academyofathens.gr}

\credit{Conceptualization, Supervision, Validation, Writing - Original Draft}

\affiliation[2]{organization={Research Center for Astronomy and Applied Mathematics, Academy of Athens},
                postcode={11527}, 
                postcodesep={}, 
                city={Athens},
                country={Greece}}

\author[1]{Georgios Alexandridis}[orcid=0000-0002-3611-8292]
\ead{gealexandri@uoa.gr}

\credit{Supervision, Validation}

\author[2]{Antonios Nathanail}[orcid=0000-0002-1655-9912]
\ead{anathanail@academyofathens.gr}

\credit{Project administration, Funding acquisition}

\cortext[cor1]{Corresponding author}


\begin{abstract}
The pulsar magnetosphere has only recently been addressed using Physics-Informed Neural Networks (PINNs), by deploying a domain-decomposition approach and treating the separatrix and equatorial current sheet as infinitesimally thin discontinuities. However, this baseline requires extensive manual hyperparameter tuning, achieves limited final accuracy and demands several hours of training. We refine this framework by introducing domain-specific neural architectures based on Kolmogorov--Arnold networks, an automated adaptive training pipeline and a physics-based convergence criterion that eliminate the need for manual calibration. The proposed methodology delivers self-consistent axisymmetric magnetosphere solutions with mean squared errors of the PDE residuals at $\mathcal{O}\left(10^{-6}\right)$ in double precision -- an improvement of two orders of magnitude over the baseline -- while achieving convergence in under 20 minutes in single precision. Importantly, the method reliably resolves stellar radii reduced by up to 80\% compared to the baseline, overcoming the severe spatial scale disparities that also challenge traditional solvers. Furthermore, by varying the flux that opens to infinity, we provide a correction to the equation that connects it to the equatorial T-point's position. The complete framework is released as the open-source library \texttt{PulsarX}.
\end{abstract}


\maketitle

\section{Introduction} \label{sec1}

Following the initial discovery of pulsars \citep{Hewish}, the foundational theoretical framework for their strongly magnetized steady-state magnetospheres was established by \citet{Goldreich}, who described a domain partitioned into a closed-line region co-rotating with the star and confined within the light cylinder, and an open-line region extending beyond it. The first solution to the pulsar magnetosphere problem was obtained by \citet{Contopoulos99} under the assumptions of axisymmetry and ideal force-free conditions, and was subsequently refined by \citet{timokhin} to establish a standard baseline for the two-dimensional problem. Building upon this foundation, a series of numerical approaches with various assumptions and approximations have since been employed to study the two- and three-dimensional pulsar magnetosphere. Force-free electrodynamics (FFE) simulations treat the plasma as a perfectly conducting fluid with negligible inertia \citep{spitkovtskyffe, contopoulosffe}, while relativistic magnetohydrodynamic (MHD) simulations account for finite plasma inertia \citep{Komissarov, Tchekhovskoy}. Time-dependent pseudo-spectral methods have been used to better understand the link between the magnetosphere and the relativistic pulsar wind \citep{Petri}, while, more recently, first-principles particle-in-cell (PIC) \citep{Cerutti, Philippov, Kalapotharakos, Bransgrove} and hybrid force-free-PIC methods have also been explored \citep{FFEPIC}.

Despite their varying assumptions, all of these approaches rely on discrete spatial grids. To circumvent the limitations associated with grid-based discretization, a recent methodological paradigm shift has introduced neural-network-based solvers, specifically within the framework of physics-informed machine learning (PIML). Pioneered by \citet{Raissi2019}, Physics-Informed Neural Networks (PINNs) act as continuous, mesh-free universal function approximators that directly embed the governing partial differential equations (PDEs), alongside relevant boundary and initial conditions or other physical constraints, into the loss function of the neural network. PINNs have demonstrated significant utility across a broad spectrum of forward and inverse problems in domains such as fluid dynamics \citep{fluids, supersonic}, solid mechanics \citep{viscoelasticity, elasticity}, and astrophysical systems \citep{teukolsky, asteroseismology}.

Within the context of the pulsar magnetosphere, the application of PINNs has been advanced by two primary research efforts. In one direction, \citet{urban} developed a generic PINN framework for astrophysical boundary value problems by embedding parameterized boundary conditions and source terms directly into the network's input space and utilized the magnetosphere of a slowly rotating neutron star as their primary case study. Following this, \citet{stefanou} extended the methodology to the more complex regime of pulsar magnetospheres. In the other direction, \citet{dimitropoulos1, dimitropoulos2, dimitropoulos3} utilized a domain decomposition approach in an attempt to establish a new baseline for the pulsar magnetosphere problem, analogous to the foundational works of \citet{Contopoulos99} and \citet{timokhin}. Their approach removed the inherent constraints of grid-based solvers, allowing them to model the separatrix and equatorial current sheets with infinitesimal thickness and treat the light cylinder without the numerical issues tied to discrete solvers. Consequently, this methodology not only reproduced previous qualitative and quantitative results, but also confirmed features under debate, such as the fact that the Y-point is actually a T-point \citep{uzdensky, contopoulos24}.

Although the framework introduced by \citet{dimitropoulos1} provides a novel approach to the pulsar problem, especially through its treatment of the equatorial sheet and the separatrix, the authors acknowledge that the methodology is subject to several limitations. Notably, the trained model exhibits limited accuracy. High computational costs are also reported, stemming from the requirement for a large number of training epochs to achieve convergence and potential implementation inefficiencies. Furthermore, the use of a standard multi-layer perceptron (MLP) architecture requires careful tuning of its parameters to provide stable solutions. The training process also requires constant supervision for the calibration of individual loss-term weights; the authors explicitly state that the process cannot proceed unmonitored and that manual intervention is mandatory to ensure physically consistent solutions. Finally, as noted in \citet{dimitropoulos2}, the adoption of a relatively large stellar radius was a constraint necessitated by the difficulty of resolving disparate spatial scales, a challenge previously documented for other numerical methods as well \citep{Petri}.

Building on the PINN-based results established for the axisymmetric pulsar magnetosphere in \citet{dimitropoulos1}, the present work focuses on refining the methodology to address the aforementioned limitations. In particular, the objective of this study is to extend these foundational results by introducing an automated, architecturally optimized solver that provides a more robust baseline for the future modeling of not only two- and three-dimensional pulsar magnetospheres, but also those of other compact astrophysical objects (e.g., black holes). Our primary contributions are summarized as follows:

\begin{itemize}
	\item We redefine the methodology within an adaptive training framework that eliminates the need for manual calibration or prior knowledge of loss-term contributions. We demonstrate that this formulation results in a more well-behaved spectrum for the associated Neural Tangent Kernel (NTK) matrix.
	
	\item We replace the standard MLP architecture with a residual-gated adaptive Kolmogorov--Arnold network (RGA KAN). We justify this transition through a comparison of geometric complexity between the two architectures.
	
	\item We run the main experiment of \citet{dimitropoulos1}, obtaining a mean squared error (MSE) of $\mathcal{O}\left(10^{-6}\right)$ for the PDE residuals in both the open- and closed-line region while reaching convergence significantly faster.
	
	\item We perform ablation studies to isolate and quantify the specific performance improvements provided by each adaptive component introduced in our framework.
	
	\item We demonstrate the capability of our solver to resolve disparate spatial scales, producing reliable results for magnetospheres with stellar radii reduced by up to 80\% compared to \citet{dimitropoulos1}.
	
	\item We derive a refined expression connecting the open flux ratio to the equatorial T-point position, yielding a high-precision correction to the standard pulsar spindown rate.
	
	\item By utilizing our \texttt{JAX}-based software, \texttt{PulsarX}, we reduce total convergence times from hours to minutes in a single GPU without sacrificing model complexity or collocation point density. We provide this tool as an open-source resource (see Reproducibility statement) to ensure reproducibility and to enable further refinement of the methodology by the community.
\end{itemize}

The remainder of this paper is organized as follows. Section \ref{sec2} introduces the pulsar magnetosphere problem in axisymmetry and the PIML framework, alongside an in-depth discussion of the current methodology's limitations. Section \ref{sec3} details our methodological advancements, providing preliminary comparisons of NTK stability and architectural geometric complexity against the existing baseline. In Section \ref{sec4} we present our experimental results, including systematic ablation studies, extensions to smaller stellar radii and results for varying open flux ratios. Finally, Section \ref{sec5} provides a summary of our findings and outlines directions for future research.

\section{Problem statement} \label{sec2}

For the mathematical formulation of the problem, we adopt a dimensionless unit system where all distances are normalized to the light cylinder radius, $R_\text{LC} \equiv c/\Omega$, where $\Omega$ is the angular velocity of stellar rotation and $c$ is the speed of light. We assume the standard force-free regime, where electromagnetic fields dominate the plasma dynamics, allowing us to neglect gravitational effects, thermal pressure and particle inertia.

\subsection{The axisymmetric pulsar equation} \label{sec2.1}

Under the aforementioned assumptions, the bulk of the magnetosphere is described by the force-free condition:

\begin{equation}
	\rho_{e} \mathbf{E} + \mathbf{J} \times \mathbf{B} = 0, \label{eq1}
\end{equation}

\noindent where $\mathbf{B}$ and $\mathbf{E}$ represent the magnetic and electric fields, respectively, and $\rho_{e}$ and $\mathbf{J}$ are the electric charge and current densities, respectively. All quantities are evaluated in a non-rotating lab frame; following the formulation of \citet{muslimov}, Faraday's law and the Ampère-Maxwell equation in this steady state translate into the following system in spherical coordinates $(r, \theta, \phi)$:


\begin{align}
	\nabla \times \left( \mathbf{E} + r \sin \theta \hat{\phi} \times \mathbf{B} \right) &= 0, \label{eq2}\\
	\nabla \times \left( \mathbf{B} - r \sin \theta \hat{\phi} \times \mathbf{E} \right) &= 4\pi \mathbf{J} - r \sin \theta \hat{\phi} \, \nabla \cdot \mathbf{E}. \label{eq3}
\end{align}

\noindent From Eq. \eqref{eq2}, it follows that the electric field is determined by the rotation of the magnetic field lines:

\begin{equation}
	\mathbf{E} = -r \sin \theta \, \hat{\phi} \times \mathbf{B} \label{eq4}
\end{equation}

\noindent By substituting Eq. \eqref{eq4} into Eq. \eqref{eq3}, the Ampère-Maxwell equation can be rewritten in terms of the poloidal (meridional) and toroidal (azimuthal) magnetic field components:

\begin{equation}
	\nabla \times \left[ \left(1 - r^2 \sin^2 \theta \right) \left(B_r \hat{r} + B_\theta\hat{\theta}\right) + B_\phi \hat{\phi} \right] = \Lambda \mathbf{B}, \label{eq5}
\end{equation}

\noindent This relation, first established by \citet{endean} and \citet{mestel}, introduces the force-free parameter $\Lambda$, defined as:

\begin{equation}
	\Lambda = \frac{1}{B^2} \left[ \mathbf{B} \cdot \left( \nabla \times \mathbf{B} \right) - (r \sin \theta) B_\phi (\nabla \cdot \mathbf{E}) \right]. \label{eq6}
\end{equation}

\noindent This parameter remains constant along individual magnetic field lines, satisfying

\begin{equation}
	\mathbf{B} \cdot \nabla \Lambda = 0. \label{eq7}
\end{equation}

In axisymmetry, i.e., $\partial \left(\cdot\right)/\partial \phi = 0$, the magnetic field is conveniently described by introducing the auxiliary scalar magnetic flux function $\Psi(r, \theta)$, which automatically satisfies $\nabla \cdot \mathbf{B} = 0$. The magnetic field components are connected to $\Psi$ via:

\begin{align}
	B_{r} &\equiv \frac{1}{r^{2} \sin \theta} \frac{\partial \Psi}{\partial \theta}, \label{eq8} \\
	B_{\theta} &\equiv -\frac{1}{r \sin \theta} \frac{\partial \Psi}{\partial r}, \label{eq9} \\
	B_{\phi} &\equiv \frac{I(\Psi)}{r \sin \theta}, \label{eq10}
\end{align}

\noindent where $I(\Psi)$ is the poloidal electric current function. The force-free parameter is related to this current function through its derivative, $\Lambda \equiv I'(\Psi)$. Substituting these components into Eq. \eqref{eq5} transforms it into the second-order partial differential equation known as the pulsar equation in axisymmetry (aligned rotator):

\begin{equation}
	(1 - r^{2} \sin^{2} \theta) \left[ \frac{\partial^{2} \Psi}{\partial r^{2}} + \frac{1}{r^{2}} \frac{\partial^{2} \Psi}{\partial \theta^{2}} - \frac{\cos \theta}{r^{2} \sin \theta} \frac{\partial \Psi}{\partial \theta} \right] - 2 \sin\theta\left( r \sin\theta \frac{\partial \Psi}{\partial r} + \cos\theta \frac{\partial \Psi}{\partial \theta} \right) + II'(\Psi) = 0. \label{eq11}
\end{equation}

\noindent We note that a physically consistent solution must remain regular across the light cylinder ($r \sin \theta = 1$), which imposes the regularization condition:

\begin{equation}
	II' = 2 \left(\sin \theta \frac{\partial \Psi}{\partial r} + \frac{\cos\theta}{r} \frac{\partial \Psi}{\partial \theta} \right), \quad \text{for} \quad r\sin\theta = R_\text{LC} = 1. \label{eq12}
\end{equation}

\noindent Interestingly, using Eqs. \eqref{eq8}--\eqref{eq10}, this can equivalently be written as

\begin{equation}
	I^\prime = \frac{2B_z}{B_\phi}, \quad \text{for} \quad r\sin\theta = R_\text{LC} = 1, \label{extraIeq}
\end{equation}

\noindent which is an equation that holds for any type of coordinate system. This condition determines the values of the current function along all field lines that cross the light cylinder and extend into the open-line region. Furthermore, the requirement of Eq. \eqref{eq7} translates into the condition that the poloidal current $I(\Psi)$ be a function of the magnetic flux alone, i.e.:

\begin{equation}
	\frac{\partial \Psi}{\partial r} \frac{\partial I}{\partial \theta} - \frac{\partial \Psi}{\partial \theta} \frac{\partial I}{\partial r} = 0. \label{eq13}
\end{equation}

\noindent Enforcing this alignment condition is critical for establishing a physically valid force-free solution.

\subsection{Physics-informed machine learning} \label{sec2.2}

To solve the axisymmetric pulsar equation defined in Eq. \eqref{eq11}, we employ the PIML framework, originally introduced by \citet{Raissi2019}. This approach reframes the problem of solving PDEs into an optimization task, where the governing physics are embedded directly into the learning process through the loss function. Consider a general steady-state PDE of the form:

\begin{equation}
	\mathcal{F}[\Psi(\mathbf{x})] = 0, \quad \mathbf{x} \in \Omega, \label{eq14}
\end{equation}

\noindent where $\mathcal{F}$ is a nonlinear differential operator and $\mathbf{x}$ denotes the spatial coordinates within the domain $\Omega$. In the PIML framework, the unknown solution $\Psi(\mathbf{x})$ is approximated by a neural network $\mathcal{N}(\mathbf{x}; \boldsymbol{\vartheta})$, where $\boldsymbol{\vartheta}$ represents the set of the network's trainable parameters. A significant advantage of this formulation is its architecture-agnostic nature; the framework imposes no constraints on the specific design of $\mathcal{N}$, allowing for the use of various architectures without altering the underlying methodology.

Deviating slightly from the original formulation of \citet{Raissi2019}, we adopt an approach where boundary conditions and other physical requirements are treated as a generalized set of constraints. We first define the PDE residual as:

\begin{equation}
	\mathcal{R}_{\text{pde}}(\mathbf{x}) \equiv \mathcal{F}[\mathcal{N}(\mathbf{x}; \boldsymbol{\vartheta})]. \label{eq15}
\end{equation}

\noindent In addition to the governing equation, we define $M$ physical constraints, with each constraint associated with a specific residual function $\mathcal{R}_i(\mathbf{x})$ for $i = 1, \dots, M$. The neural network is then trained to minimize the total loss function, $\mathcal{L}$, defined as the weighted sum of the MSE of these residuals:

\begin{equation}
	\mathcal{L}(\boldsymbol{\vartheta}) = \lambda_{\text{pde}} \mathcal{L}_{\text{pde}}(\boldsymbol{\vartheta}) + \sum_{i=1}^{M}{ \lambda_i \mathcal{L}_i(\boldsymbol{\vartheta}) }, \label{eq16}
\end{equation}

\noindent where $\lambda_{\text{pde}}$ and $\lambda_i$ are the global weight coefficients for the PDE and the $i$-th physical constraint, respectively. The individual loss terms are computed over distinct sets of collocation points:

\begin{equation}
	\mathcal{L}_{\text{pde}}(\boldsymbol{\vartheta}) = \frac{1}{N_{\text{pde}}} \sum_{j=1}^{N_{\text{pde}}} \left| \mathcal{R}_{\text{pde}}\left(\mathbf{x}_\text{pde,j}\right) \right|^2, \qquad
	\mathcal{L}_i(\boldsymbol{\vartheta}) = \frac{1}{N_i} \sum_{j=1}^{N_i} \left| \mathcal{R}_i \left(\mathbf{x}_{i,j}\right) \right|^2, \label{eq17}
\end{equation}

\noindent where $N_{\text{pde}}$ is the number of points sampled within the domain to impose the PDE, and $N_i$ is the number of collocation points utilized to impose the $i$-th physical constraint. While this formulation serves as the default PIML setup, it can be modified to incorporate additional adaptive techniques to improve convergence in complex physical scenarios, as will be discussed in Section \ref{sec3}.

\subsection{Existing methodology \& limitations} \label{sec2.3}

In the work of \citet{dimitropoulos1}, the pulsar magnetosphere is addressed using a domain decomposition approach with two individual PINNs trained: one to predict the magnetic flux, $\Psi_\text{cl}$, in the closed-line region, where $I(\Psi) = 0$, and another to predict both the flux, $\Psi_\text{op}$, and the current function, $I_\text{op}$, in the open-line region. These two domains are separated by the separatrix surface, which is modeled by a third neural network. The objective of this third network is to learn the shape of the surface, which in axisymmetry can be defined as a single-valued function of the polar angle $\theta$, $r_s(\theta)$. Because the separatrix is a mathematical contact discontinuity that PINNs cannot automatically resolve, the methodology involves an iterative adjustment process.

Starting from an arbitrary initial shape (typically the dipole configuration), the separatrix is incrementally deformed until the pressure balance condition $(B^2 - E^2)_{r = r_s^-} = (B^2 - E^2)_{r = r_s^+}$ is satisfied across the boundary. During each adjustment cycle, the two PINNs are first trained within their respective domains. Subsequently, the magnitudes of the electric and magnetic fields are evaluated at a selected set of polar angles along the current separatrix. For each angle $\theta_i$, the radial position $r_s\left(\theta_i\right)$ is updated according to:

\begin{equation}
	r_s\left(\theta_i\right) \to r_s\left(\theta_i\right) + \beta P_\text{ratio} \left(\frac{r_s\left(\theta_i\right)-R_*}{1-R_*}\right)^2 , \label{eq18}
\end{equation}

\noindent where $R_*$ is the stellar radius, $\beta$ is a relaxation parameter and $P_\text{ratio}$ is a pressure ratio defined as:

\begin{equation}
	P_\text{ratio} = 2 \, \frac{(B^2 - E^2)_{r = r_s^-} - (B^2 - E^2)_{r = r_s^+}}{(B^2 - E^2)_{r = r_s^-} + (B^2 - E^2)_{r = r_s^+}}. \label{eq19}
\end{equation}

\noindent The adjusted radial values at these discrete polar angles are then utilized as training targets to re-train the separatrix neural network in a supervised setting. The authors report that 10 cycles of 50,000 training iterations each are generally required to reach a physical solution when using MLPs as the underlying architecture for their PINNs (see Section 4.5 of \citet{dimitropoulos1}).

In order to handle the equatorial current sheet, which is the other primary source of discontinuity, the authors employ a field-reversal technique originally proposed by \citet{bogovalov}. This ``trick'' effectively makes the equatorial current sheet disappear during the training phase, allowing it to be reactivated once convergence is achieved. Central to this formalism is the control of the total magnetic flux that opens to infinity, denoted as $\Psi_S$. This value is determined by the prescribed opening angle of the polar cap, $\theta_{\text{pc}}$, according to the relation $\Psi_S \equiv \Psi_{\max} \sin^2\theta_{\text{pc}}$, where $\Psi_{\max}$ is the maximum flux at the stellar surface. Under this setup, $\Psi = 0$ is imposed as a boundary condition on the positive $z$-axis, while $\Psi = 2\Psi_S$ is set on the negative $z$-axis to satisfy the field-reversal requirement. Additionally, $\Psi = \Psi_S$ is enforced on both the separatrix and the equator, and a standard dipole configuration, $\Psi = \Psi_{\max} \sin^2\theta$, is imposed on the stellar surface $r = R_*$.

To be able to impose Dirichlet conditions as $r \to \infty$, the authors introduce the coordinate transformations $\mu = \cos\theta$ and $q = 1/r$, so that $r \to \infty$ corresponds to $q = 0$. Furthermore, they solve for the auxiliary fields $f$ and $\mathcal{I}$, where $\Psi\left(r,\theta\right) = \sin^2\theta \, f\left(r,\theta\right)$ and $I\left(\Psi\right) = \sin^2\theta \,\mathcal{I}\left(\Psi\right)$. While these fields are smoother, they result in significantly more complex forms of Eq. \eqref{eq11}, thus increasing the computational cost of evaluating the PDE residuals. Critically, the definition for the $f$ field inherently fails to satisfy $\Psi = 2\Psi_S$ on the negative $z$-axis ($\theta = \pi$), as the $\sin^2\theta$ term vanishes. Consequently, the authors are restricted to working only on the positive $z$-plane, thus needing to impose additional symmetry conditions on the equator to recover the global solution.

Apart from the challenges introduced by field transformations, the framework faces additional limitations. Specifically, the authors report that the enforcement of the alignment condition of Eq. \eqref{eq13} requires a high number of training iterations within each adjustment cycle, which significantly extends the overall computational time. When combined with potential implementation inefficiencies, this leads to training for several hours on a single GPU to reach convergence. Furthermore, the process involves constant manual monitoring and calibration. On one hand, the reliance on standard MLP architectures not only limits predictive performance compared to more advanced architectures but also introduces significant depth-scaling issues within the PINN setting \citep{piratenets}; this, in turn, leads to the requirement for careful manual tuning to identify appropriate values for network width and depth. On the other hand, the presence of competing loss terms is a well-documented challenge in the training of PINNs \citep{wang21}. Manually calibrating the global weight coefficients ($\lambda_{\text{pde}}, \lambda_i$) represents a non-trivial task, especially in problems where the number of simultaneous physical constraints is high; consequently, algorithms that dynamically adapt the values of these weights during training are typically favored \citep{minmax, wangntk}. Finally, the existing methodology is constrained by the scale of the domain. Current results are limited to a relatively large stellar radius ($R_* = 0.25$), as PINNs often struggle to resolve disparate spatial scales \citep{dimitropoulos2} -- a challenge not exclusive to PINNs, as it has also been noted in other approaches \citep{Petri}. Even for such high values of $R_*$, the accuracy achieved corresponds to a MSE of the PDE residuals on the order of $\mathcal{O}\left(10^{-4}\right)$ in the open-line region and even higher in the closed-line region.

\section{Proposed improvements} \label{sec3}

To address the identified limitations, the methodology proposed in this work is built upon three main pillars: a high-performance re-implementation of the computational framework, the adoption of a more expressive neural architecture adapted to the needs of the specific problem, as well as the introduction of an adaptive training pipeline that eliminates the need for manual tuning. Regarding the first pillar, we implemented the existing framework from scratch using \texttt{JAX} \citep{jax} and refined it with the adaptive components presented herein. \texttt{JAX} has emerged as one of the most prominent libraries for PIML due to its support for just-in-time (jit) compilation and vectorization (\texttt{vmap}), which significantly accelerate the computation of high-order derivatives and the evaluation of large batches of collocation points. To support the scientific community and encourage further research, we provide the entire codebase as an open-source GitHub repository titled \texttt{PulsarX} (see Reproducibility statement). This repository includes extensive documentation and is designed to serve as a robust baseline for studying the magnetospheres of pulsars and other compact objects. The architectural innovations and the adaptive training strategies constituting the other two pillars are presented in detail in the following subsections.

\subsection{Neural architectures} \label{sec3.1}

As established, standard MLP architectures suffer from limited expressivity and scale poorly with increasing depth, thus requiring careful manual tuning to identify optimal width and depth. PirateNets \citep{piratenets} represent the current state-of-the-art for MLP-based PINNs, offering improved scaling stability with increasing number of parameters while mitigating the training divergences of plain MLPs. However, the recently introduced residual-gated adaptive Kolmogorov--Arnold networks (RGA KANs) offer similar scaling advantages while demonstrating superior accuracy across various PDE benchmarks \citep{rga}. RGA KANs use Kolmogorov--Arnold networks (KANs) \citep{kan1, kan2} as their backbone, which exhibit reduced spectral bias compared to MLPs \citep{spectral} and have proven highly effective in several PIML applications \citep{adaptivekan, fair, pinnreview}. For the purposes of this work, we implement the RGA KAN architecture within \texttt{PulsarX} using the \texttt{jaxKAN} library \citep{jaxkan}; further details regarding the structure of RGA KANs are provided in Appendix \ref{appA}.

In time-dependent PDE problems, the output layer of an RGA KAN is often pre-trained to fit the initial condition, providing a physically motivated starting point for training. While a similar strategy for the pulsar magnetosphere could involve pre-training the networks to approximate a standard dipole configuration,\footnote{Preliminary numerical experiments indicated that this pre-training approach yielded significantly less satisfactory results compared to the use of functional ansätze.} we instead embed physical structure directly through problem-specific ansätze. Rather than solving for the auxiliary $f$ and $\mathcal{I}$ fields in the transformed system of \citet{dimitropoulos1}, we retain the original form of Eq. \eqref{eq11} and define two RGA KAN models: $\mathcal{N}_{\text{cl}}(r, \theta)$ with a scalar output for the closed-line region, and $\mathcal{N}_{\text{op}}(r, \theta)$ with a two-dimensional output $\mathcal{N}_{\text{op}, 1}, \mathcal{N}_{\text{op}, 2}$, for the open-line region. The physical fields are then approximated as:

\begin{align}
	\Psi_{\text{cl}}(r, \theta) &\approx \sin^2\theta \, \mathcal{N}_{\text{cl}}(r, \theta), \label{eq20} \\
	\Psi_{\text{op}}(r, \theta) &\approx (1 - \cos\theta)\Psi_S + \sin^2\theta \, \mathcal{N}_{\text{op}, 1}(r, \theta), \label{eq21} \\
	I_{\text{op}}(r, \theta) &\approx \sin^2\theta \, \mathcal{N}_{\text{op}, 2}(r, \theta). \label{eq22}
\end{align}

\noindent This formulation exploits the fact that the network output need not coincide with the physical fields. The ansätze enforce the $z$‑axis boundary conditions as hard constraints rather than penalty terms in the loss function; one readily verifies that $\Psi_{\text{op}}\left(r,0\right) = 0$ (positive $z$-axis) and $\Psi_{\text{op}}\left(r,\pi\right) = 2\Psi_S$ (negative $z$-axis). Furthermore, we adopt the $\sin^2\theta$ pre-factor used by \citet{dimitropoulos1} in their definition of the auxiliary $f$ and $\mathcal{I}$ fields, but without switching to a description based on these fields. This approach maintains consistency with the field reversal trick while eliminating the need for additional soft constraints in the loss function.

In addition to the two PINNs, we define a third neural network, $\mathcal{N}_{\text{sep}}(\theta)$, dedicated to the supervised learning of the adjusted separatrix radii at the end of each training cycle. As this task involves mapping a one-dimensional input to a one-dimensional output along a continuous curve, the full capacity of an RGA KAN is unnecessary. We therefore utilize a ModifiedMLP architecture \citep{wang21} augmented with Random Fourier Feature (RFF) embeddings \citep{rff} to mitigate the spectral bias typical of standard MLPs. To take advantage of the problem's axisymmetry, the separatrix function is parametrised as:

\begin{equation}
	r_s(\theta) \approx \mathcal{N}_{\text{sep}}(\theta^2). \label{eq23}
\end{equation}

\noindent Our investigation indicates that while this specific ansatz is not critical for the system to converge to a physical solution, it can significantly help the training process by reducing the total number of separatrix adjustment cycles required.

\subsection{Adaptive training framework} \label{sec3.2}

Following the domain decomposition strategy, the composite loss function $\mathcal{L}$ that is optimized to train the aforementioned neural networks is the weighted sum of ten individual components:

\begin{align}
	\mathcal{L}\left(\boldsymbol{\vartheta}\right) = 
	\,& \lambda_{\text{pde}}^{\text{op}} \mathcal{L}_{\text{pde}}^{\text{op}}\left(\boldsymbol{\vartheta}\right) + \lambda_{\text{pde}}^{\text{cl}} \mathcal{L}_{\text{pde}}^{\text{cl}}\left(\boldsymbol{\vartheta}\right) + 
	\lambda_{\text{align}}^{\text{op}} \mathcal{L}_{\text{align}}^{\text{op}}\left(\boldsymbol{\vartheta}\right) +
	\lambda_{\text{star}}^{\text{op}} \mathcal{L}_{\text{star}}^{\text{op}}\left(\boldsymbol{\vartheta}\right) + \lambda_{\text{star}}^{\text{cl}} \mathcal{L}_{\text{star}}^{\text{cl}}\left(\boldsymbol{\vartheta}\right) + \nonumber \\
	&\lambda_{\text{sep}}^{\text{op}} \mathcal{L}_{\text{sep}}^{\text{op}}\left(\boldsymbol{\vartheta}\right) + \lambda_{\text{sep}}^{\text{cl}} \mathcal{L}_{\text{sep}}^{\text{cl}}\left(\boldsymbol{\vartheta}\right) +
	\lambda_{\text{eq}}^{\text{op}} \mathcal{L}_{\text{eq}}^{\text{op}}\left(\boldsymbol{\vartheta}\right) +
	\lambda_{\text{far}}^{\text{op}} \mathcal{L}_{\text{far}}^{\text{op}}\left(\boldsymbol{\vartheta}\right) + \lambda_{\text{lc}}^{\text{op}} \mathcal{L}_{\text{lc}}^{\text{op}}\left(\boldsymbol{\vartheta}\right), \label{eq24}
\end{align}

\noindent where each term's superscript denotes the regional PINN (open-line or closed-line) that is used to calculate the loss, while the subscript indicates the specific physical constraint being imposed. Within this sum, the first two terms enforce the PDE residuals of Eq. \eqref{eq11}, while the third term enforces the alignment condition of Eq. \eqref{eq13}. The next two terms enforce the stellar surface condition at $r = R_*$, which is defined piecewise to apply the field-reversal technique:

\begin{equation}
	\Psi = \Psi_{\max} \cdot
	\begin{cases}
		 \sin^2 \theta, & \text{for} \quad 0 \le \theta \le \pi/2, \\
		2 - \sin^2 \theta, & \text{for} \quad \pi/2 < \theta \le \pi.
	\end{cases}
\end{equation}

\noindent The following three terms impose $\Psi = \Psi_S$ on the separatrix boundary defined by $\mathcal{N}_{\text{sep}}$, and the equator, while the ninth term imposes a Neumann boundary condition, $\partial \Psi / \partial r = 0$, at the outer domain boundary $R_{\text{max}}$. Since we work in spherical coordinates rather than the transformed $q=1/r$ space, this condition allows field lines to exit the simulation box smoothly by assuming that at a sufficiently large distance, the poloidal component of the magnetic field becomes radial. Finally, the tenth term enforces the light cylinder regularization condition of Eq. \eqref{eq12}. Although this condition is theoretically satisfied if the PDE is solved accurately, explicitly penalizing the residual at $r\sin\theta=1$ provides an early gradient signal that steers the open-line-region PINN toward the correct current function $I(\Psi)$, avoiding the long convergence times reported by \citet{dimitropoulos1} for the alignment condition.

Manually balancing the competing gradients from these ten loss components would require extensive trial and error. Instead, we automate this process by implementing a learning rate annealing (LRA) algorithm \citep{wang21} that dynamically adjusts the global weights $\lambda_i$ during training. Every few iterations, the gradient magnitudes of each loss term with respect to the parameters $\boldsymbol{\vartheta}$ are computed and temporary weights are formed as:

\begin{equation}
	\hat{\lambda}_i = \frac{\sum_{j=1}^{10} \lVert\nabla_{\boldsymbol{\vartheta}} \mathcal{L}_j(\boldsymbol{\vartheta}) \rVert_2}{\lVert\nabla_{\boldsymbol{\vartheta}} \mathcal{L}_i(\boldsymbol{\vartheta}) \rVert_2}, \label{eq25}
\end{equation}

\noindent where $\lVert \cdot \rVert_2$ denotes the $L^2$ norm. The global weights are then updated using the rule:

\begin{equation}
	\lambda_i \to a \lambda_i + (1-a)\hat{\lambda}_i, \label{eq26}
\end{equation}

\noindent where $a$ is a mixing factor, typically set to $0.95$, and all weights are initialized at $1$. This procedure ensures that if a specific loss term has high gradients that threaten to dominate the total loss, the other terms are scaled up accordingly to remain relevant in the parameter update.

While these global weights balance the different physical laws, we also apply Residual-Based Attention (RBA) \citep{rba} to resolve spatial imbalances within each individual loss term. This method introduces local weights $\alpha_{i,j}$ for each collocation point, effectively redefining the individual loss terms as:

\begin{equation}
	\mathcal{L}_i(\boldsymbol{\vartheta}) = \frac{1}{N_i} \sum_{j=1}^{N_i} \left| \alpha_{i,j} \, \mathcal{R}_i \left(\mathbf{x}_{i,j}\right) \right|^2, \label{eq27}
\end{equation}

\noindent where $\mathcal{R}_i\left(\mathbf{x}_{i,j}\right)$ is the residual at the $j$-th collocation point of the $i$-th loss term. These local weights are initialized at $1$ and updated at every training iteration according to:

\begin{equation}
	\alpha_{i,j} \to \gamma\alpha_{i,j} + \eta \frac{ \left| \mathcal{R}_i \left(\mathbf{x}_{i,j}\right) \right| }{ \max_k \left(\left\{ \left| \mathcal{R}_i \left(\mathbf{x}_{i,k}\right) \right| \right\}_{k=1}^{N_i} \right) }, \label{eq28}
\end{equation}

\noindent where $\gamma, \eta$ are the method's hyperparameters, typically set to $\gamma = 0.999$ and $\eta = 0.01$. To ensure the networks generalize well and avoid overfitting, we periodically resample the collocation points. Since this resampling occurs much less frequently than the RBA weight updates, we re-initialize the local weights to $1$ at the start of each resampling event to allow the attention mechanism to re-evaluate the new distribution. While alternative strategies could involve maintaining a large pool of persistent weights and performing adaptive sampling \citep{rga}, our investigation showed that such methods primarily increased the training time per cycle without providing a significant boost to the final accuracy of the solution.

Beyond these adaptive weighting techniques, we introduce some refinements to the iterative framework itself, beginning with a modified update rule for the separatrix surface. In place of the adjustment rule of Eq. \eqref{eq18}, we propose updating the separatrix radii that are used as targets for the training of $\mathcal{N}_\text{sep}$ as follows:

\begin{equation}
	r_s\left(\theta_i\right) \to r_s\left(\theta_i\right) + \beta P_\text{ratio} \left[ 1.0 - \left( \frac{|\theta_i - \pi/2|}{\pi/2 - \theta_{\text{pc}}} \right)^{k} \right], \label{eq29}
\end{equation}

\noindent where $k$ is a large exponent, typically set to $k=12$. The newly introduced term is designed to enforce the physical constraint that the separatrix remains anchored to the stellar surface at the polar caps, as the update term vanishes at $\theta = \theta_{\text{pc}}$. The high exponent $k$ effectively introduces a step function behavior, where the term inside the parentheses remains close to zero across mid-latitudes and becomes exactly zero at the equator ($\theta = \pi/2$). We find that this rule prevents the separatrix from unphysically detaching from the star at high latitudes and significantly accelerates the convergence towards its physical shape.

Finally, we replace the empirical approach of a fixed number of training cycles with a physics-based termination criterion. We define convergence based on two simultaneous conditions, which are not independent of each other: the geometric stability of the separatrix and the achievement of pressure equilibrium across the boundary. Specifically, the iterative process of separatrix adjustment terminates when the mean radial shift of the separatrix across all polar angles falls below a tolerance $\epsilon_s$, and the maximum pressure ratio $P_\text{ratio}$ at any point on the surface is lower than a threshold $\epsilon_P$. The choice of $P_\text{ratio}$ as a convergence metric is a deliberate move to normalize the pressure signal across the domain. Due to the dipolar nature of the magnetic field, the pressure near the stellar surface is higher than near the light cylinder. Consequently, an absolute pressure difference criterion would be biased toward the polar caps and could potentially compromise accuracy at the equatorial T-point.

\subsection{Preliminary comparative study} \label{sec3.3}

To perform a preliminary verification on the advantages of our framework, we analyze the optimization dynamics and model expressivity through two theoretical diagnostics: the NTK \citep{ntk} and the geometric complexity \citep{complexity}. Although the NTK framework was originally developed to analyze standard MLPs in the infinite-width limit, it has since been extended to the PIML framework \citep{wangntk} and recently utilized in KAN-based architectures to evaluate different architectural and optimization-based aspects of training \citep{kanntk, kaninit}. Geometric complexity, measured via a discrete Dirichlet energy over the domain, provides a quantitative measure of the spatial variability of a network's predictions. Within the PIML setting, geometric complexity has been used to compare the convergence dynamics and overfitting tendencies between competing architectures \citep{kkan}.

To this end, we conduct two experiments. We adopt the benchmark configuration used by \citet{dimitropoulos1}, choosing a stellar radius of $R_* = 0.25$, a polar cap opening angle of $\theta_{\text{pc}} = 1.176\sqrt{R_*}$ and a maximum surface flux of $\Psi_{\max} = 1.0$. Training is performed on a NVIDIA RTX 4090 GPU with single precision. Rather than running the full, multi-cycle iterative process, we initialize the separatrix to a dipole shape and train each setup for 50,000 iterations, which corresponds to a single training cycle in the baseline work.

The first setup serves as the baseline, utilizing standard MLP architectures with the exact parameters of \citet{dimitropoulos1}: a network consisting of 2 input neurons, 3 hidden layers of 64 neurons each and 1 output neuron for the closed-line region, alongside a matching configuration with 2 output neurons for the open-line region. In this baseline case, RBA and LRA are deactivated and collocation points are resampled every 2,000 iterations. The second case represents our adaptive framework, employing RGA KANs for the two PINNs. To ensure a fair comparison, we explicitly constrain the capacity of the RGA KANs to match the parameter number of the baseline MLPs; specifically, we implement a single-block RGA KAN with a hidden dimension of 20 for both regions, yielding 8,640 parameters for $\mathcal{N}_\text{op}$ (compared to 8,642 for the MLP) and 8,539 parameters for $\mathcal{N}_\text{cl}$ (compared to 8,577 for the MLP); further details regarding the architecture are provided in Appendix \ref{appA}. Additionally, the neural ansätze defined in Eqs. \eqref{eq20}--\eqref{eq22} are enforced identically across both the MLPs and RGA KANs to isolate the effect of the underlying network architecture. In this adaptive case, RBA is activated with default hyperparameters ($\gamma = 0.999$, $\eta = 0.01$) and LRA is performed every 250 iterations with a mixing factor of $a = 0.95$.

\begin{figure}[t!]
	\centering
	\includegraphics[width=\textwidth]{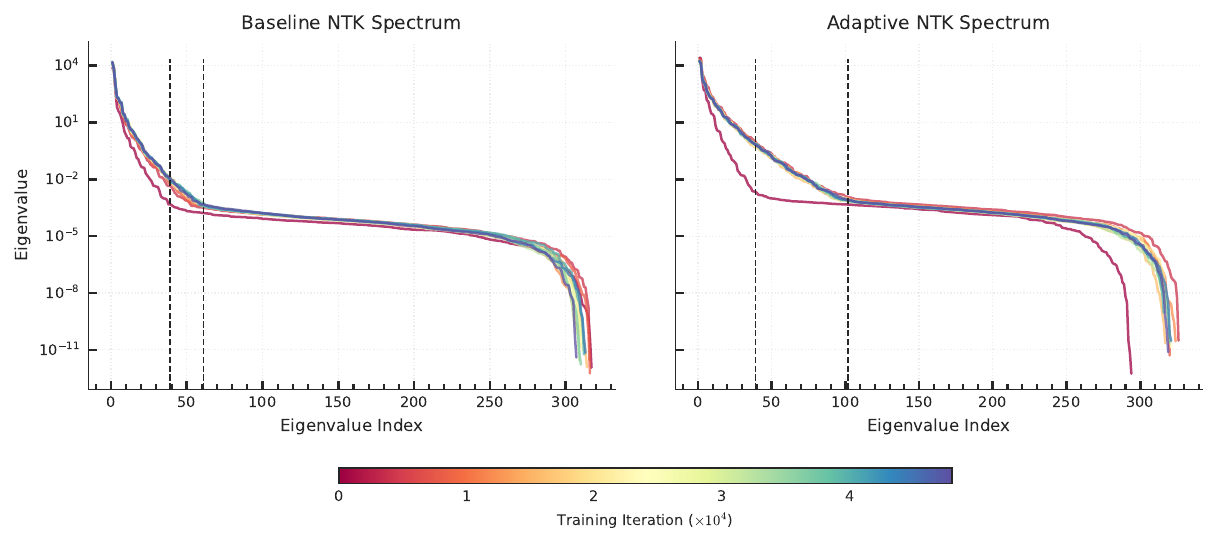}
	\caption{Evolution of the NTK eigenvalue spectra during 50,000 training iterations for the baseline MLP configuration (left) and the proposed adaptive framework (right). The color scale indicates the training progression ($\times 10^4$ iterations), moving from initialization (dark red) to the final iteration (blue). Vertical dashed lines indicate the expansion of the non-plateau eigenvalue regime.}
	\label{fig1}
\end{figure}

\begin{figure}[b!]
\centering
\includegraphics[width=\textwidth]{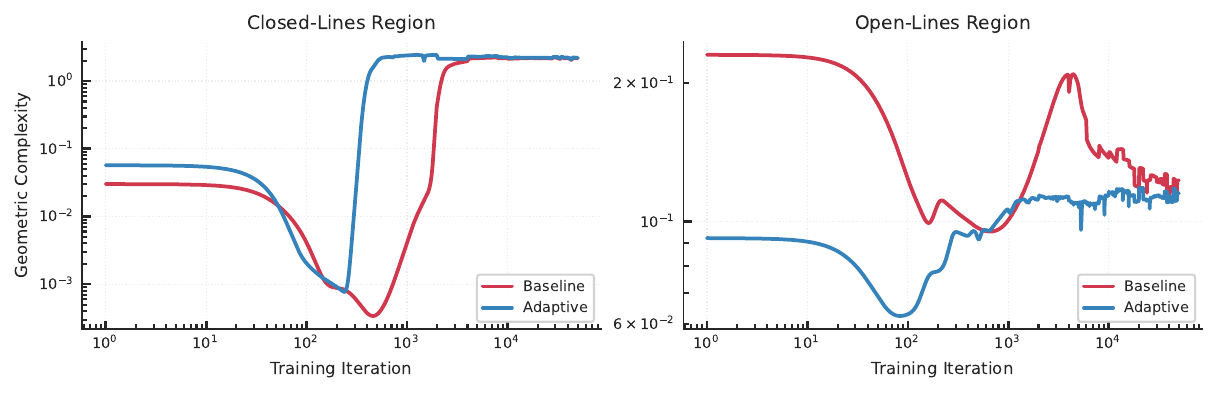}
\caption{Evolution of the geometric complexity during the first 50,000 training iterations for the closed-line region network $\mathcal{N}_\text{cl}$ (left) and the open-line region network $\mathcal{N}_\text{op}$ (right). Both panels compare the standard MLP baseline (red) against the RGA KAN architecture of the adaptive framework (blue).}
\label{fig2}
\end{figure}

In both experiments, identical collocation point counts are used to enforce all 10 physical constraints: $2^{10}$ points for the PDE in the closed-line region, $2^{11}$ points for the PDE and alignment condition in the open-line region, $2^8$ points per region for the stellar surface, $2^8$ points per region along the separatrix boundary, $2^9$ points for the equatorial condition, $2^9$ points for the outer boundary and $2^8$ points for the light cylinder regularization condition. Optimization is performed via a joint Adam optimizer \citep{adam} initialized with a learning rate of $5 \cdot 10^{-4}$, featuring a linear warmup period of 1,000 iterations followed by an exponential decay at a rate of 0.9 every 2,000 iterations. Throughout the training process, we record the geometric complexity of the models at every iteration, and compute the eigenvalues of the NTK every 4000 iterations, using a representative subsample of $2^8$ points per region. The results are shown in Fig. \ref{fig1} and Fig. \ref{fig2} for the NTK spectra and geometric complexities, respectively.

As illustrated in Fig. \ref{fig1}, the NTK eigenvalue spectra for both the baseline and adaptive configurations exhibit a qualitatively similar profile characterized by a rapid power-law decay that flattens into a plateau, before finally dropping to practically zero. However, in the baseline configuration, the spectrum evolves slowly and continuously throughout the entire training trajectory. At initialization (dark red curve), the transition to the plateau occurs near the 40th eigenvalue index (indicated by the first vertical dashed line). As training progresses, this boundary shifts marginally to around index 60 (indicated by the second vertical dashed line), lifting a small subset of eigenvalues out of the plateau. In contrast, the adaptive framework exhibits a rapid and much more pronounced evolution. Within the earliest phase of training, the spectrum abruptly shifts to the right, moving the plateau boundary from index 40 to approximately index 100. Following this sharp initial transition, the subsequent eigenvalue curves bundle tightly together, demonstrating that the kernel quickly reaches a stable configuration. We note that the minor discrepancies in the maximum eigenvalue index where the curves terminate on the far right are purely artifacts of numerical precision; because the NTK spectrum spans from $\mathcal{O}\left(10^4\right)$ down to $\mathcal{O}\left(10^{-12}\right)$, eigenvalues approaching the lower bound can evaluate as slightly negative due to floating-point precision and are consequently discarded. This overall behavior highlights two key advantages of our adaptive refinements to the framework. First, the rapid settling of the NTK spectrum confirms that LRA and RBA quickly equilibrate the gradient contributions from the ten loss terms, preventing any single term from dominating the parameter updates. Second, the significant expansion of the non-plateau region, containing more than double the number of ``elevated'' eigenvalues compared to the baseline, indicates that a substantially larger subspace of feature directions is being actively optimized.

In Fig. \ref{fig2}, both architectures share the same qualitative evolutionary profile: a short initial phase of constant complexity, followed by a transient decrease (compression), a subsequent sharp rise and an ultimate convergence to a stable plateau. This multi-phase complexity trajectory is consistent with prior theoretical analyses of KAN-based architectures \citep{kkan}. The critical difference lies in the convergence timescale. In the closed-line region, the RGA KAN completes its transition within a few hundred iterations, whereas the MLP undergoes a prolonged expansion phase that lasts for several thousand iterations. This disparity is even more pronounced in the open-line region, where the RGA KAN smoothly approaches its plateau around the 1,000th iteration, exhibiting only minor, localized variations thereafter. In contrast, the MLP undergoes a significant complexity increase, peaking near the 4,000th iteration, and does not converge until well beyond 10,000 iterations. These observations directly validate the limitation noted by \citet{dimitropoulos1}: the original framework required large training cycles spanning 50,000 iterations to allow the intermediate models to stabilize before adjusting the separatrix. By replacing the MLPs with RGA KANs and incorporating RBA and LRA into the optimization, we drastically compress this inner-loop convergence timescale.

\section{Experimental results} \label{sec4}

Following the preliminary single-cycle analysis, we implement the full multi-cycle iterative procedure to resolve the self-consistent separatrix geometry. The geometric complexity results from Section \ref{sec3.3} demonstrate that the networks in both regions reach their structural plateaus significantly earlier compared to the MLP baseline. We may therefore adopt a significantly more conservative number of iterations per cycle than the baseline framework, so we train the regional networks for 10,000 iterations per cycle before executing each adjustment of the separatrix boundary.

\subsection{Main benchmark} \label{sec4.1}

We begin by replicating the exact physical experiment presented in \citet{dimitropoulos1} prior to evaluating alternative configurations of the physical system, so we use the same values for $R_*$, $\theta_{\text{pc}}$ and $\Psi_{\max}$ as in Section \ref{sec3.3}. All parameters of the adaptive training pipeline are also set to the values used in that section, with some minor exceptions. Since a direct parameter-matching with the baseline MLP is no longer required to ensure a fair architectural comparison, we reduce the hidden dimension of the RGA block in $\mathcal{N}_{\text{cl}}$ to 15, yielding 4,909 parameters to account for the smaller spatial volume of the closed-line region. Additionally, collocation points are resampled every 5,000 iterations, corresponding to the initiation and midpoint of each cycle, while LRA is executed every 2,000 iterations.

The separatrix boundary is parameterized by a ModifiedMLP, $\mathcal{N}_\text{sep}$, with 2 hidden layers of 64 neurons each (16,994 parameters). While reducing the hidden dimension (e.g., to 32 or 16) yields practically identical results, we opt for a number of parameters matching that of the baseline work's network, since $\mathcal{N}_\text{sep}$ is updated only once per cycle and thus introduces negligible computational overhead. The network is initialized to a dipole shape and optimized independently via Adam with an initial learning rate of $5 \cdot 10^{-3}$ and an exponential weight decay rate of 0.6 applied every 2,000 iterations. At the end of each regional training cycle, $\mathcal{N}_\text{sep}$ undergoes supervised training for 2,000 iterations using a dataset evaluated at 1,000 equidistant polar angles.

Global convergence of the iterative process is declared when the geometric stability threshold $\epsilon_s = 0.0015$ and the pressure equilibrium threshold $\epsilon_P = 0.1$ are simultaneously satisfied. The two orders of magnitude disparity between these thresholds accounts for the relaxation parameter $\beta = 0.025$ in the update rule of Eq. \eqref{eq18}. To ensure statistical significance, all evaluations are performed across 3 independent random seeds and executed in both single precision (FP32) and double precision (FP64).

\begin{figure}[t!]
	\centering
	\includegraphics[width=0.9\textwidth]{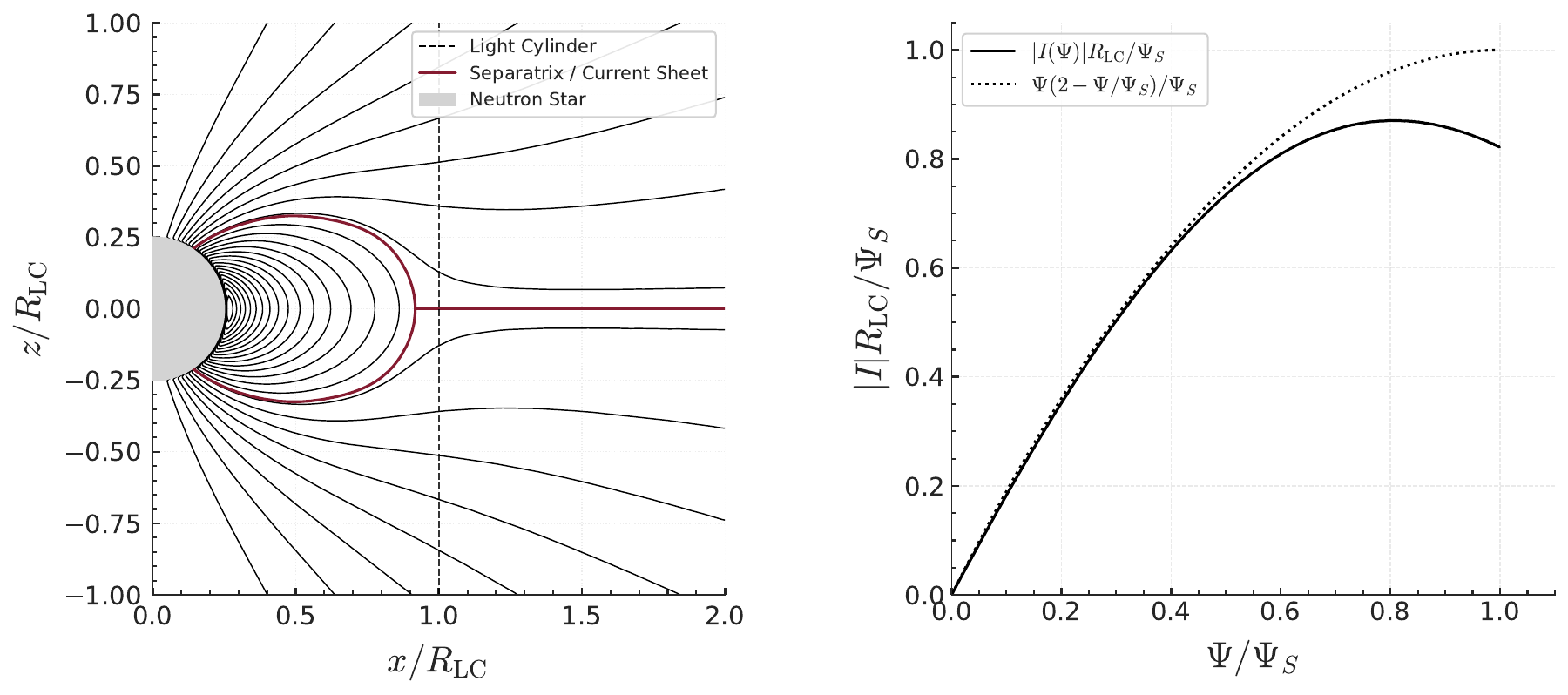}
	\caption{Self-consistent pulsar magnetosphere solution under single precision (FP32), where $z = r\cos\theta$ and $x = r\sin\theta$. Left: Magnetic field lines shown as contours of constant flux. The equatorial T-point is located at $x_T = 0.919 \, R_{\text{LC}}$. Right: Normalized poloidal current distribution $|I(\Psi)| R_{\text{LC}} / \Psi_S$ across the open magnetic field lines. The dotted line denotes the theoretical analytic expression for the split monopole.}
	\label{fig3}
\end{figure}

Across the three independent random seeds evaluated in single precision, the framework reached convergence in 13 cycles for two of the runs and 18 cycles for the third. Upon convergence, the MSE of the PDE residuals is $(6.89 \pm 3.48) \cdot 10^{-5}$ for the closed-line region and $(3.40 \pm 0.84) \cdot 10^{-5}$ for the open-line region, where we report the $\text{mean} \pm \text{standard error}$. The equatorial point of the separatrix is located at $0.919 \pm 0.002 \, R_{\text{LC}}$, a position slightly further outward than the value of $0.88 \, R_{\text{LC}}$ reported by \citet{dimitropoulos1}. The self-consistent solution for this baseline configuration is presented in Fig. \ref{fig3}, which shows the magnetic field lines as iso-contours of the flux (left plot) and the converged poloidal current (right plot). Importantly, the methodology recovers a distinct T-point geometry at the equator rather than a Y-point configuration, matching the critical qualitative structure originally predicted by \citet{uzdensky} and confirmed in \citet{dimitropoulos1}.

Excluding the initial cycle, which requires an average of $4.12$ minutes due to jit compilation overhead, the framework achieves an average runtime of $0.997 \pm 0.031$ minutes per cycle. This corresponds to an execution speed of $5.98$ ms per training iteration, factoring in the concurrent optimization of the regional PINNs, the application of the adaptive training algorithms and the supervised updates of the separatrix boundary. Given an average convergence requirement of $14.66$ cycles, a complete self-consistent solution is obtained in approximately $17.7$ minutes. Compared to the multiple hours required by the baseline methodology or the days demanded by traditional grid-based numerical solvers \citep{dimitropoulos1}, this represents a major computational speedup achieved alongside significantly lower final PDE residuals.

\begin{figure}[t!]
	\centering
	\includegraphics[width=\textwidth]{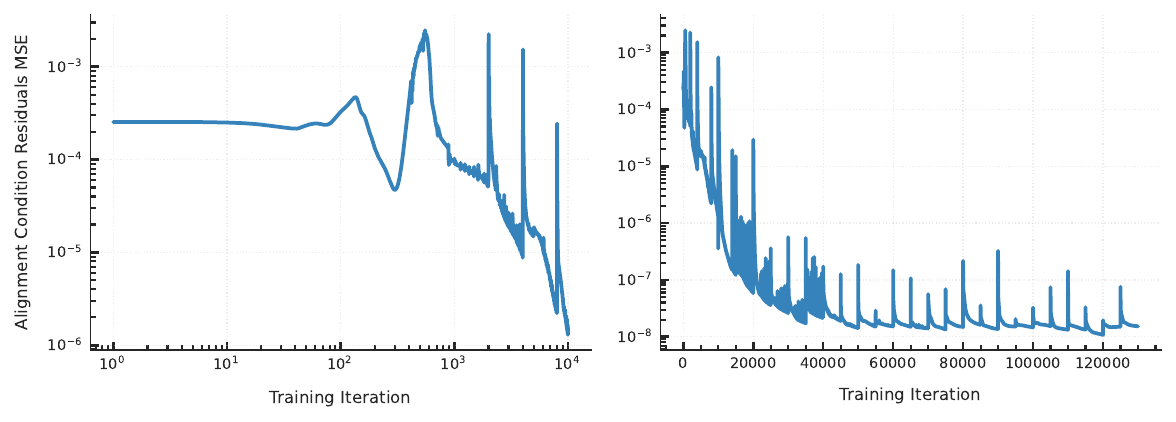}
	\caption{Evolution of the MSE of the alignment condition's residuals during single-precision training. Left: Close-up of the first training cycle (initial 10,000 iterations) plotted on a logarithmic horizontal axis. Right: Complete multi-cycle optimization trajectory spanning all training cycles.}
	\label{fig4}
\end{figure}

To confirm that our approach deals with another limitation of the baseline framework, Fig. \ref{fig4} tracks the MSE of the alignment condition's residuals for the run that converged after 18 cycles. As shown in the localized first-cycle view (Fig. \ref{fig4}, left), the network actively begins minimizing the alignment constraint around the 1,000th iteration. This directly coincides with the timeframe where the open-line-region network's geometric complexity starts stabilizing into its plateau (see Fig. \ref{fig2}). By the end of the first cycle, the alignment residual falls to $\mathcal{O}\left(10^{-6}\right)$. This is a massive acceleration in optimization compared to the baseline framework of \citet{dimitropoulos1}, where the slow convergence of this specific condition was cited as a primary reason for requiring 50,000-iteration cycles.

\begin{figure}[b!]
\centering
\includegraphics[width=0.9\textwidth]{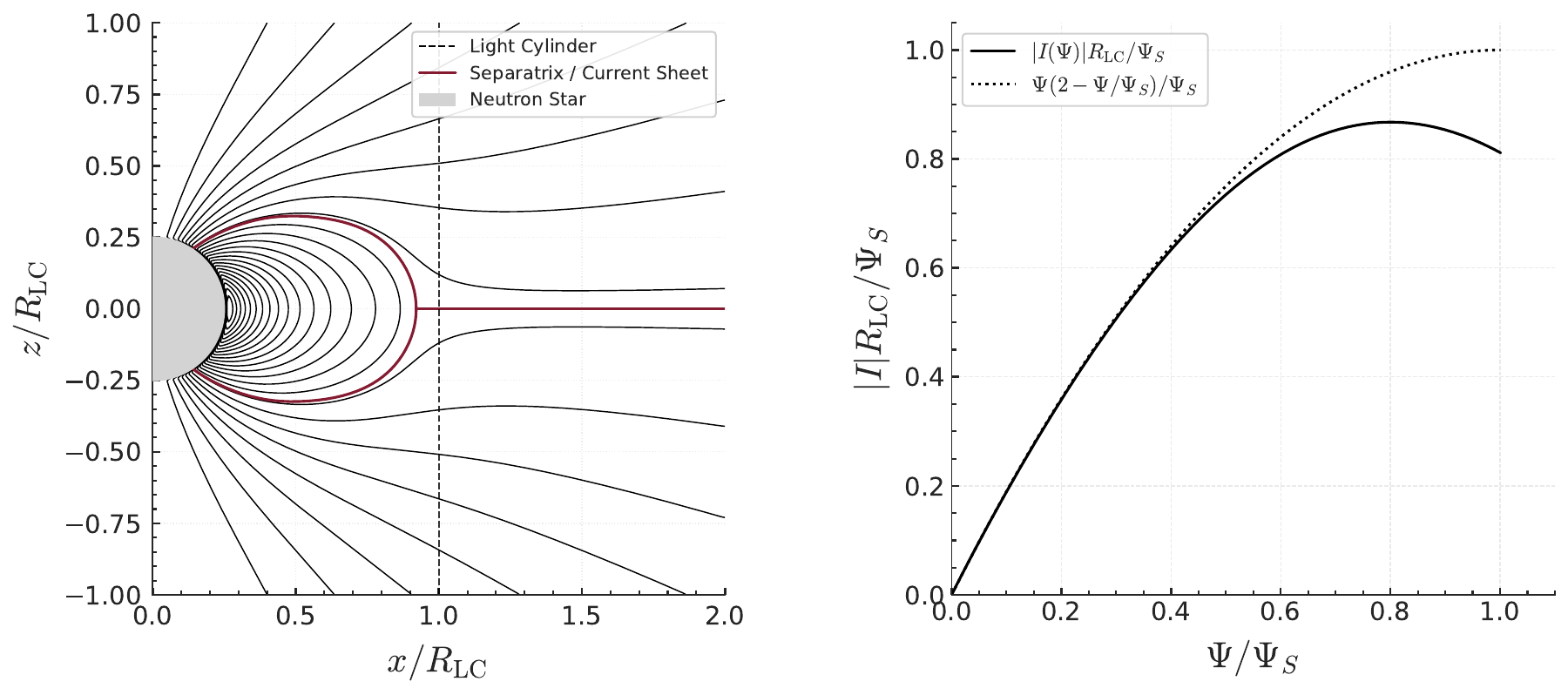}
\caption{Same as Figure \ref{fig3}, under double precision (FP64). The equatorial T-point is located at $x_T = 0.925 \, R_{\text{LC}}$.}
\label{fig5}
\end{figure}

Over the full multi-cycle process (Fig. \ref{fig4}, right), the alignment loss continues its downward progression, ultimately reaching a minimum $\mathcal{O}\left(10^{-8}\right)$. The periodic transient spikes correspond exactly to the scheduled execution of LRA. Importantly, the execution of geometric separatrix updates at the 10,000-iteration cycle boundaries does not induce severe loss spikes and force the model to re-minimize the relevant loss term from scratch, contrasting sharply with the behavior reported in Fig. 3 of \citet{dimitropoulos1}. 

Executing the same benchmark in double precision incurs a higher computational penalty, increasing the average runtime per cycle to $2.579 \pm 0.003$ minutes. Under this configuration, convergence is achieved in 14, 15 and 17 cycles across the respective seeds. However, this increased time cost yields an order-of-magnitude reduction in errors, with the final PDE residual MSE dropping to $(4.06 \pm 1.78) \cdot 10^{-6}$ for the closed-line region and $(6.46 \pm 2.37) \cdot 10^{-6}$ for the open-line region. In this case, the T-point is resolved on average at $0.925 \pm 0.002 \, R_{\text{LC}}$, shifting slightly closer to the light cylinder than predicted in the single-precision runs. Because low final accuracy was explicitly identified as a structural bottleneck in the original framework, these FP64 results offer a rigorous numerical correction to the value reported in \citet{dimitropoulos1}. The corresponding fields are illustrated in Fig. \ref{fig5}; the global profiles remain qualitatively identical to those in Fig. \ref{fig3}, with the minor exception of the separatrix extending marginally further outward due to the improved numerical accuracy.

Finally, one could claim that the observed accuracy gains over the baseline framework stem primarily from solving the problem on a finite domain in spherical coordinates rather than the compactified space of the ($q, \mu$) coordinates used in the original work. For this reason, we evaluate our methodology under double precision within the transformed coordinate system in Appendix \ref{appB}. The final residual errors are slightly higher than these reported for the spherical-coordinate system; however, they remain significantly lower than our single-precision spherical results and substantially below those reported by \citet{dimitropoulos1}. Additionally, the T-point location is resolved at $0.922 \pm 0.001 \, R_{\text{LC}}$, which is in close statistical agreement with the spherical double-precision result, confirming that our improvements cannot be mainly attributed to the underlying coordinate representation.

\subsection{Ablation study} \label{sec4.2}

To isolate and quantify the individual contributions of our adaptive framework's components, we perform an ablation study across four configurations:

\begin{itemize}
	\item \textbf{Ablation 1 (Architecture):} All neural networks are reverted to standard MLPs. To keep constant parameter numbers compared to the RGA KAN/ModifedMLP setup, $\mathcal{N}_\text{op}$ is configured with 3 hidden layers of 64 neurons, $\mathcal{N}_\text{cl}$ with 2 hidden layers of 64 neurons, and $\mathcal{N}_\text{sep}$ with 2 hidden layers of 128 neurons.
	\item \textbf{Ablation 2 (No RBA):} The RGA KAN architecture and LRA are kept intact, but RBA is deactivated.
	\item \textbf{Ablation 3 (No LRA):} The RGA KAN architecture and RBA are kept intact, but LRA is deactivated.
	\item \textbf{Ablation 4 (No Ansätze):} The ansätze defined in Eqs. \eqref{eq20}--\eqref{eq22} are completely removed. The boundary conditions along the $z$ axis are instead enforced explicitly by introducing two additional loss constraints (one for $z>0$ and one for $z<0$) evaluated over $2^9$ collocation points.
\end{itemize}

All ablation experiments are executed in single precision across 3 independent random seeds. A failsafe termination criterion of 35 maximum cycles is imposed; a cycle count of 35 indicates that the framework failed to satisfy the convergence thresholds ($\epsilon_s, \epsilon_P$). For successful runs, the solution is deemed ``physical'' if it accurately recovers the core qualitative field distributions shown in Fig. \ref{fig3} and successfully resolves the equatorial T-point. Finally, for runs that are both convergent and physically valid, we report the final MSE of the regional PDE residuals. The comparative results are summarized in Table \ref{tab1}.

\begin{table}[width=.9\linewidth,cols=7,pos=h]
	\caption{Quantitative summary of the ablation study across three independent random seeds (FP32).}
	\label{tab1}
	\begin{tabular*}{\tblwidth}{@{} CCCCCCC@{} }
		\toprule
		\textbf{Ablation} & \textbf{Seed} & \textbf{Cycles} & \textbf{Physical} & $\mathbf{x_T \, \left(R_{\text{LC}}\right)}$ & \textbf{Closed-line PDE MSE} & \textbf{Open-line PDE MSE} \\
		\midrule
		 & 1 & 35 & — & — & — & — \\
		1 & 2 & 35 & — & — & — & — \\
		 & 3 & 35 & — & — & — & — \\
		 \midrule
		 & 1 & 16 & Yes & 0.925 & $1.62 \cdot 10^{-4}$ & $1.12 \cdot 10^{-5}$ \\
		2 & 2 & 16 & Yes & 0.921 & $1.27 \cdot 10^{-4}$ & $7.90 \cdot 10^{-5}$ \\
		 & 3 & 17 & Yes & 0.921 & $1.46 \cdot 10^{-4}$ & $3.15 \cdot 10^{-5}$ \\
		 \midrule
		 & 1 & 16 & Yes & 0.916 & $2.18 \cdot 10^{-5}$ & $1.82 \cdot 10^{-6}$ \\
		3 & 2 & 35 & — & — & — & — \\
		 & 3 & 35 & — & — & — & — \\
		 \midrule
		 & 1 & 18 & No & 0.603 & — & — \\
		4 & 2 & 17 & No & 0.489 & — & — \\
		 & 3 & 35 & — & — & — & — \\
		 \bottomrule
	\end{tabular*}
\end{table}

Regarding the first ablation, reverting to standard MLPs prevents the framework from achieving convergence within the 35-cycle limit across all seeds. This indicates that under the specific physical convergence thresholds used herein, conventional MLP architectures are inadequate unless alternative parameters (requiring manual tuning) or architectural enhancements (e.g., RFF embeddings) are used, or the thresholds are significantly relaxed. On the other hand, deactivating RBA for the second ablation still leads to physical convergence across all evaluated seeds, requiring an average of $16.3$ cycles and yielding a mean T-point location of $x_T = 0.922 \pm 0.001 \, R_{\text{LC}}$. The final averaged MSE values for PDE residuals are $(1.45 \pm 0.10) \cdot 10^{-4}$ for the closed-line region and $(4.06 \pm 2.01) \cdot 10^{-5}$ for the open-line region. These results reveal that, while RBA is not mandatory to achieve convergence, removing it degrades the closed-line-region accuracy by an entire order of magnitude and increases the average computational cost by 2 cycles.

\begin{figure}[t!]
	\centering
	\includegraphics[width=0.9\textwidth]{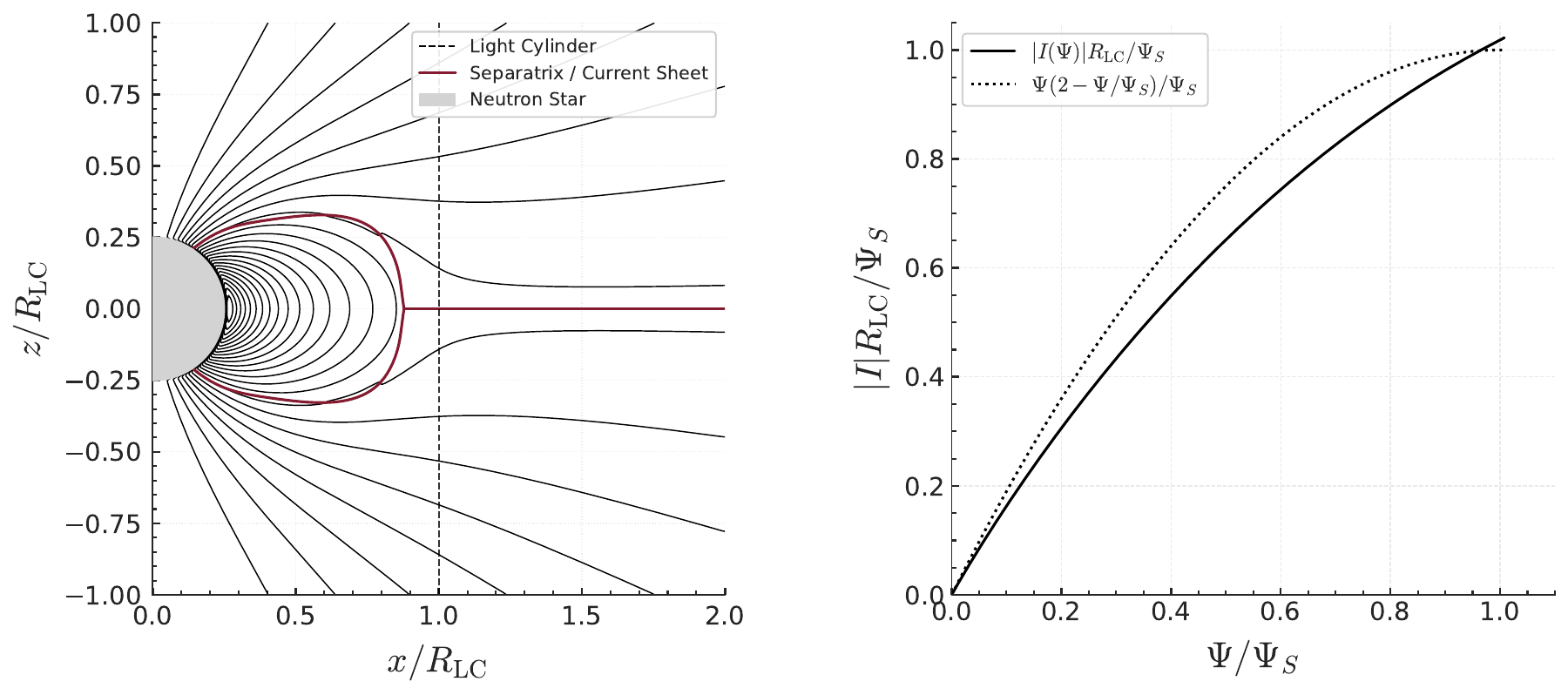}
	\caption{Unphysical magnetosphere solution obtained when both RBA and LRA are deactivated under single precision (FP32), where $z = r\cos\theta$ and $x = r\sin\theta$. Left: Magnetic field lines shown as contours of constant flux. Right: Normalized poloidal current distribution $|I(\Psi)| R_{\text{LC}} / \Psi_S$ across the open magnetic field lines.}
	\label{fig6}
\end{figure}

Turning to the third ablation, the impact of deactivating LRA is more severe, with only 1 out of 3 runs reaching convergence. Interestingly, this single successful run produces the lowest single-precision MSE achieved in the study ($2.18 \cdot 10^{-5}$ for the closed-line region and $1.82 \cdot 10^{-6}$ for the open-line region) with a slightly displaced T-point at $x_T = 0.916 \, R_{\text{LC}}$. While such an isolated success could be an anomaly of weight initialization, it is primarily explained by the fact that RBA remains active. In the absence of LRA's global balancing, RBA's local weight adaptation can occasionally regulate the competing losses well enough to guide a favorable seed to convergence. To verify this hypothesis, we executed three additional runs with both RBA and LRA simultaneously deactivated and found only a single converging run. However, this run corresponded to a heavily distorted, unphysical magnetosphere, illustrated in Fig. \ref{fig6}. Specifically, the resolved separatrix is geometrically deformed, with the equatorial T-point protruding slightly outward. Furthermore, the extrema of the poloidal current distribution are elevated in absolute value. Most critically, the magnetic field line topology suffers from unphysical artifacts; there exist field lines originating from the stellar surface within the open-line region which penetrate directly into the separatrix boundary before re-emerging and exiting it downstream. This confirms that RBA cannot independently substitute LRA, which is foundational for reproducible convergence, while RBA acts as a secondary mechanism to maximize accuracy.

Finally, removing the ansätze of Eqs. \eqref{eq20}--\eqref{eq22} for the purposes of the fourth ablation also leads to failure. While two of the runs achieve convergence under the set thresholds, the resulting magnetospheres are entirely unphysical. As indicated by the severely misplaced T-point locations ($x_T = 0.6028 \, R_{\text{LC}}$ and $x_T = 0.4886 \, R_{\text{LC}}$), the open-line region practically suppresses the closed-line one, forcing the separatrix boundary inward. In summary, the ablation study results demonstrate that reliable and automated extraction of the axisymmetric pulsar magnetosphere requires the combined deployment of all proposed adaptive components, with the exception of RBA which simply increases the converged results' accuracy.

\subsection{Results for smaller stars} \label{sec4.3}

\begin{figure}[b!]
	\centering
	\includegraphics[width=\textwidth]{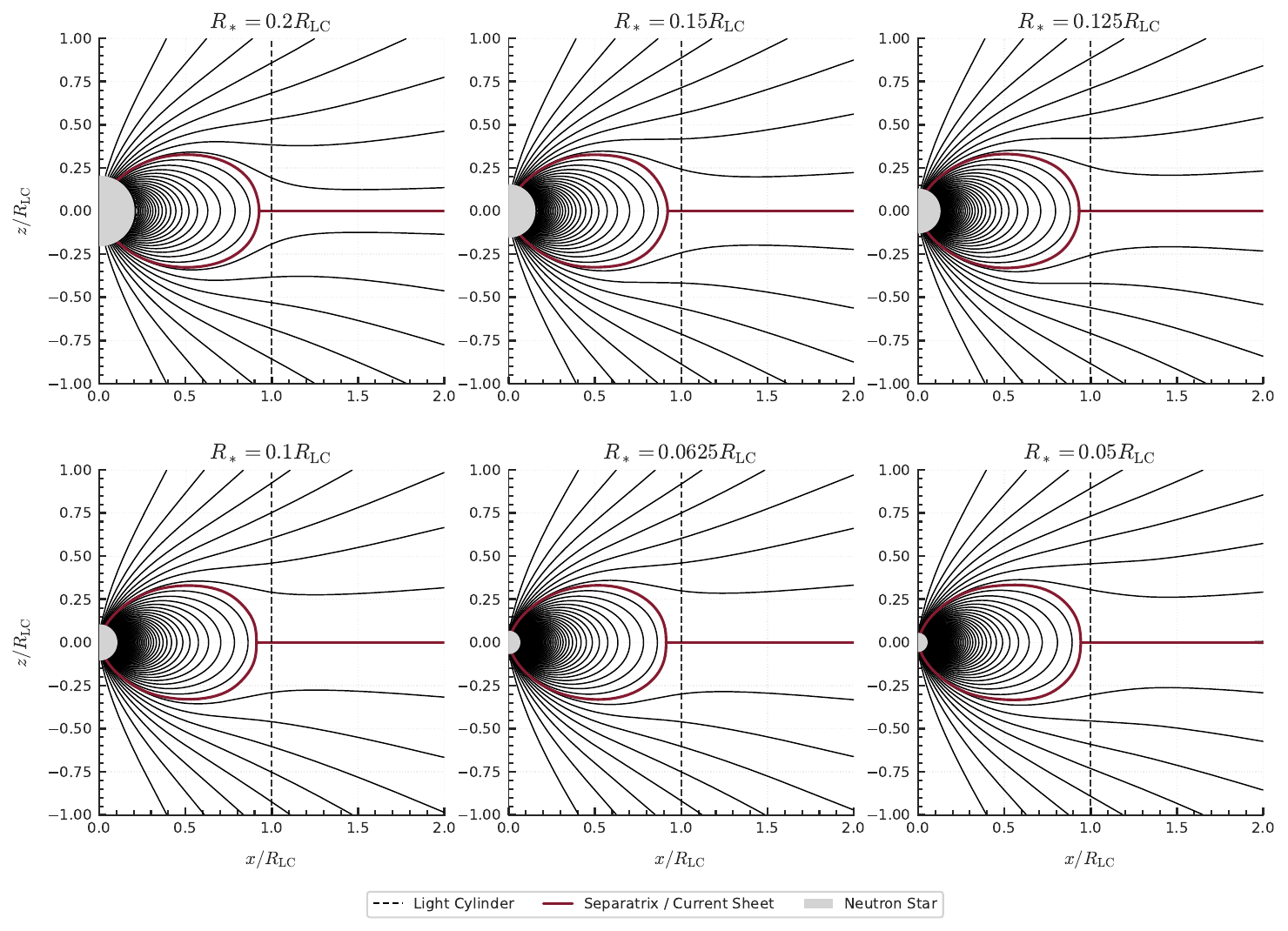}
	\caption{Converged magnetic field line configurations for decreasing stellar radii at polar cap angle $\theta_\text{pc} = \arcsin\sqrt{1.231R_*}$: $R_* = 0.2$ (top-left), $R_* = 0.15$ (top-middle), $R_* = 0.125$ (top-right), $R_* = 0.1$ (bottom-left), $R_* = 0.0625$ (bottom-middle), and $R_* = 0.05$ (bottom-right). Across all configurations, the maximum deviation of the equatorial T-point from the baseline location ($x_T = 0.925 \, R_{\text{LC}}$) is $0.009 \, R_{\text{LC}}$.}
	\label{fig7}
\end{figure}

To assess the effectiveness of the framework beyond the standard benchmark configuration ($R_* = 0.25$), we extend our analysis to smaller stellar radii. We systematically reduce the stellar radius across six configurations: $R_* = 0.2, 0.15, 0.125, 0.1, 0.0625$, and $0.05$, corresponding to a reduction of up to 80\% from the baseline. The polar cap opening angle $\theta_{\text{pc}}$ is scaled for each configuration to maintain a constant ratio between the magnetic flux that opens to infinity and the ideal dipole flux at the light cylinder:

\begin{equation}
	\rho = \frac{\Psi_S}{\Psi_{\text{dipole LC}}}, \label{eq30}
\end{equation}

\noindent where $\Psi_{\text{dipole LC}} = \Psi_{\max} R_*$. For $\Psi_\text{max} = 1.0$ and $\rho \approx 1.231$, the polar cap angle is parameterized as $\theta_\text{pc} = \arcsin\sqrt{1.231 R_*}$. Under this condition, and if the size of the star does not influence the final solution, we expect the equatorial T-point $x_T$ to remain unchanged. All evaluations are performed across three independent random seeds in double precision, due to the expanded spatial range. This increased spatial scale disparity also requires training the regional networks for 20,000 iterations during the first six adjustment cycles before reverting to the standard 10,000 iterations per cycle. The averaged quantitative performance metrics are reported in Table \ref{tab2} and the corresponding magnetic field lines of the converged solutions are presented in Fig. \ref{fig7}.

\begin{table}[width=.9\linewidth,cols=5,pos=h]
	\caption{Quantitative summary for decreasing stellar radii under double precision (FP64). Values represent the mean $\pm$ standard error across 3 independent seeds, with the baseline $R_* = 0.25$ run included as a reference.}
	\label{tab2}
	\begin{tabular*}{\tblwidth}{@{} CCCCC@{} }
		\toprule
		$\mathbf{R_* \, \left(R_{\text{LC}}\right)}$ & \textbf{$\boldsymbol{\theta}_{\text{pc}} \, (\text{rad})$} & $\mathbf{x_T \, \left(R_{\text{LC}}\right)}$ & \textbf{Closed-line PDE MSE} & \textbf{Open-line PDE MSE} \\
		\midrule
		0.25 & 0.588 & $0.925 \pm 0.002$ & $(4.07 \pm 1.78) \cdot 10^{-6}$ & $(6.46 \pm 2.37) \cdot 10^{-6}$ \\
		0.20 & 0.519 & $0.927 \pm 0.001$ & $(7.48 \pm 2.70) \cdot 10^{-6}$ & $(1.74 \pm 0.97) \cdot 10^{-5}$ \\
		0.15 & 0.444 & $0.926 \pm 0.003$ & $(4.69 \pm 1.02) \cdot 10^{-5}$ & $(2.01 \pm 0.48) \cdot 10^{-5}$ \\
		0.125 & 0.403 & $0.920 \pm 0.005$ & $(1.23 \pm 0.54) \cdot 10^{-4}$ & $(5.64 \pm 3.01) \cdot 10^{-5}$ \\
		0.10 & 0.358 & $0.916 \pm 0.003$ & $(1.22 \pm 0.24) \cdot 10^{-4}$ & $(3.89 \pm 2.54) \cdot 10^{-5}$ \\
		0.0625 & 0.281 & $0.919 \pm 0.002$ & $(2.92 \pm 0.77) \cdot 10^{-4}$ & $(1.66 \pm 1.00) \cdot 10^{-5}$ \\
		0.05 & 0.251 & $0.925 \pm 0.009$ & $(1.21 \pm 0.47) \cdot 10^{-3}$ & $(1.59 \pm 0.24) \cdot 10^{-4}$ \\
		\bottomrule
	\end{tabular*}
\end{table}

As shown in Table \ref{tab2}, the T-point remains practically consistent across all configurations. The maximum deviation from the baseline value is $0.009 \, R_{\text{LC}}$, maintaining third-decimal accuracy and empirically confirming the prediction that the position of the T-point is invariant under the constant flux ratio condition. Furthermore, the magnetic field lines presented in Fig. \ref{fig7} confirm that all resolved configurations are physically valid and free of numerical artifacts, demonstrating that the framework reliably handles the transition to smaller stellar sizes. We note, however, that the severe spatial scale disparity introduced by shrinking the star leads to a progressive increase in the final errors. The average MSE of the PDE residuals systematically climbs from $\mathcal{O}\left(10^{-6}\right)$ at the baseline $R_* = 0.25$ up to $\mathcal{O}\left(10^{-3}\right)$ for the closed-line region and $\mathcal{O}\left(10^{-4}\right)$ for the open-line region at $R_* = 0.05$. This radius defines the practical threshold of our current setup, which is why the radius was not decreased beyond this value. 

To achieve accurate convergence at even smaller physical scales, additional strategies would require implementation on top of our proposed framework. One potential approach is hierarchical domain decomposition, which would introduce an additional regional PINN and an internal artificial boundary within the closed-line domain. Another viable option is curriculum training \citep{curriculum}, where the PINNs are not trained from scratch for each configuration; instead, optimization begins at a well-resolved baseline (e.g., $R_* = 0.25$) and the stellar radius is incrementally decreased, using the pre-trained weights of the converged models as a warm-start initialization for successive steps. Nevertheless, the independent 80\% reduction in stellar radius achieved by our standalone framework represents a significant advancement.

\subsection{Results for varying flux ratios} \label{sec4.4}

The confirmation that the equatorial T-point's location is invariant to the stellar radius under a constant flux ratio carries an important practical implication: it justifies performing any calculation of the pulsar magnetosphere using a computationally favorable stellar radius, where the framework achieves its highest accuracy. Specifically, we can fix the radius at the well-resolved baseline of $R_* = 0.25$ and investigate the effect of varying the polar cap opening angle without keeping the flux ratio, $\rho$, constant. Our primary objective is to search for a flux ratio $\rho$ that pushes the T-point even closer to the light cylinder limit ($x_T \approx R_{\text{LC}} \equiv 1$), as the baseline ratio of $\rho = 1.231$ placed it at $x_T = 0.925 \, R_{\text{LC}}$. 

\begin{figure}[b!]
\centering
\includegraphics[width=\textwidth]{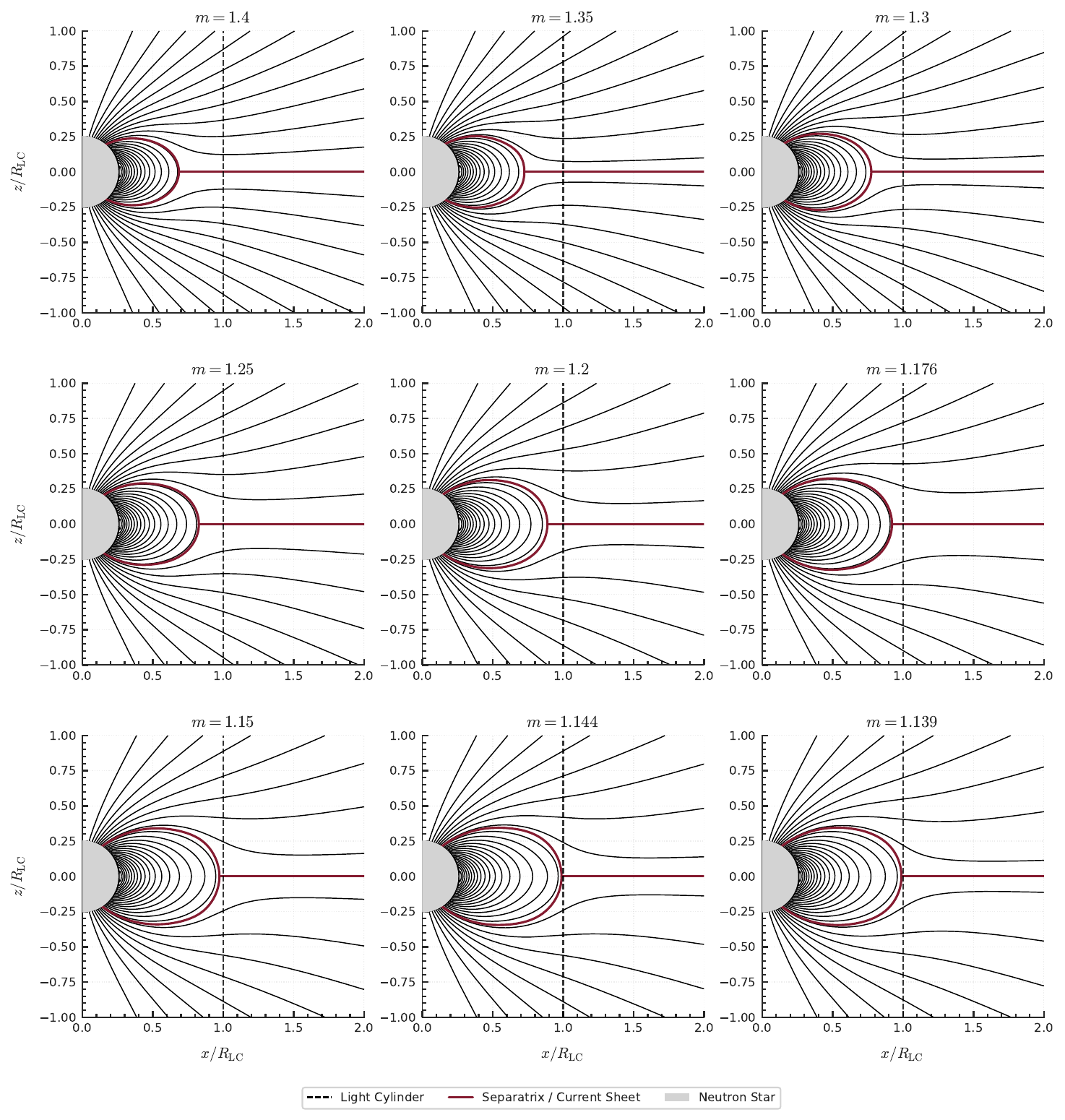}
\caption{Converged magnetic field line configurations for $R_* = 0.25$ and varying flux ratios at polar cap opening angle $\theta_\text{pc} = m\sqrt{R_*}$: $m = 1.4$ (top-left), $m = 1.35$ (top-middle), $m = 1.3$ (top-right), $m = 1.25$ (center-left), $m = 1.2$ (center-middle), $m = 1.176$ (center-right), $m = 1.15$ (bottom-left), $m = 1.144$ (bottom-middle), and $m = 1.139$ (bottom-right).}
\label{fig8}
\end{figure}

To this end, we generalize the previous polar cap angle relation to $\theta_{\text{pc}} = m\sqrt{R_*}$, where $m$ is a tunable multiplier. The standard baseline configuration corresponds to $m = 1.176$. By systematically decreasing $m$, we reduce the amount of magnetic flux that opens to infinity, which should theoretically force the T-point to expand outward. In addition to testing smaller multipliers, we also run the experiment for values of $m > 1.176$ to capture the magnetosphere under higher flux ratios. In this case, the product $\rho \cdot x_T$ should remain constant across different values of $m$, even if the individual values of $\rho$ and $x_T$ vary \citep{dimitropoulos2}. As in previous sections, we execute all runs across three independent random seeds using double precision to maintain consistency. The aggregated quantitative metrics are summarized in Table \ref{tab3} and the corresponding converged magnetic field line configurations are displayed in Fig. \ref{fig8}.

\begin{table}[width=.9\linewidth,cols=6,pos=h]
	\caption{Quantitative summary for $R_* = 0.25$ and varying flux ratios under double precision (FP64). Values represent the mean $\pm$ standard error across 3 independent seeds. The baseline run with $m = 1.176$ is highlighted in bold.}
	\label{tab3}
	\begin{tabular*}{\tblwidth}{@{} CCCCCC@{} }
		\toprule
		$\mathbf{m}$ & \textbf{$\boldsymbol{\theta}_{\text{pc}} \, (\text{rad})$} & $\boldsymbol{\rho}$ & $\mathbf{x_T \, \left(R_{\text{LC}}\right)}$ & \textbf{Closed-line PDE MSE} & \textbf{Open-line PDE MSE} \\
		\midrule
		1.400 & 0.700 & 1.660 & $0.685 \pm 0.001$ & $(6.97 \pm 2.26) \cdot 10^{-7}$ & $(2.25 \pm 0.91) \cdot 10^{-6}$ \\
		1.350 & 0.675 & 1.562 & $0.725 \pm 0.001$ & $(9.11 \pm 1.62) \cdot 10^{-7}$ & $(2.70 \pm 1.10) \cdot 10^{-6}$ \\
		1.300 & 0.650 & 1.465 & $0.773 \pm 0.001$ & $(2.21 \pm 1.36) \cdot 10^{-6}$ & $(1.48 \pm 0.73) \cdot 10^{-6}$ \\
		1.250 & 0.625 & 1.369 & $0.826 \pm 0.001$ & $(2.94 \pm 1.27) \cdot 10^{-6}$ & $(3.15 \pm 1.64) \cdot 10^{-6}$ \\
		1.200 & 0.600 & 1.275 & $0.890 \pm 0.001$ & $(4.42 \pm 0.53) \cdot 10^{-6}$ & $(1.78 \pm 1.28) \cdot 10^{-5}$ \\
		\textbf{1.176} & \textbf{0.588} & \textbf{1.231} & $\mathbf{0.925 \pm 0.002}$ & $\mathbf{(4.07 \pm 1.78) \cdot 10^{-6}}$ & $\mathbf{(6.46 \pm 2.37) \cdot 10^{-6}}$ \\
		1.150 & 0.575 & 1.183 & $0.971 \pm 0.001$ & $(2.86 \pm 0.71) \cdot 10^{-4}$ & $(2.47 \pm 0.43) \cdot 10^{-5}$ \\
		1.144 & 0.572 & 1.172 & $0.984 \pm 0.002$ & $(2.37 \pm 0.81) \cdot 10^{-4}$ & $(3.67 \pm 1.15) \cdot 10^{-5}$ \\
		1.139 & 0.570 & 1.163 & $0.988 \pm 0.002$ & $(6.20 \pm 1.22) \cdot 10^{-4}$ & $(2.01 \pm 0.98) \cdot 10^{-5}$ \\
		\bottomrule
	\end{tabular*}
\end{table}

As demonstrated by the magnetic field lines in Fig. \ref{fig8}, the framework consistently converges to physically valid magnetospheres. For larger values of the multiplier $m$ (and, by extension, higher open flux ratios $\rho$), the equatorial T-point retreats significantly inward from the light cylinder. In these configurations, the numerical accuracy of the method is exceptionally high; the MSE of the PDE residuals drops to $\mathcal{O}(10^{-7})$ in the closed-line region and $\mathcal{O}(10^{-6})$ in the open-line region, with the cross-seed variance in the T-point location remaining practically negligible. Conversely, as the flux ratio decreases, the amount of open magnetic flux is reduced, forcing the closed-line region to expand and pushing the T-point toward the light cylinder limit. As $x_T \to R_{\text{LC}}$, the accuracy of the method naturally degrades, though it remains physically satisfactory. This drop in precision is an expected consequence of the infinitesimally thin separatrix approaching the singular boundary of the light cylinder. Despite this, we identify a limiting configuration at $m = 1.139$ (corresponding to $\theta_{\text{pc}} = 0.57$ rad), which yields $x_T = 0.988 \pm 0.002 \, R_{\text{LC}}$. This represents the absolute closest approximation for $x_T \to R_\text{LC}$ achievable under this methodology's precision, providing a significant correction to standard numerical baselines \citep{timokhin}.

\begin{figure}[t!]
	\centering
	\includegraphics[width=0.5\textwidth]{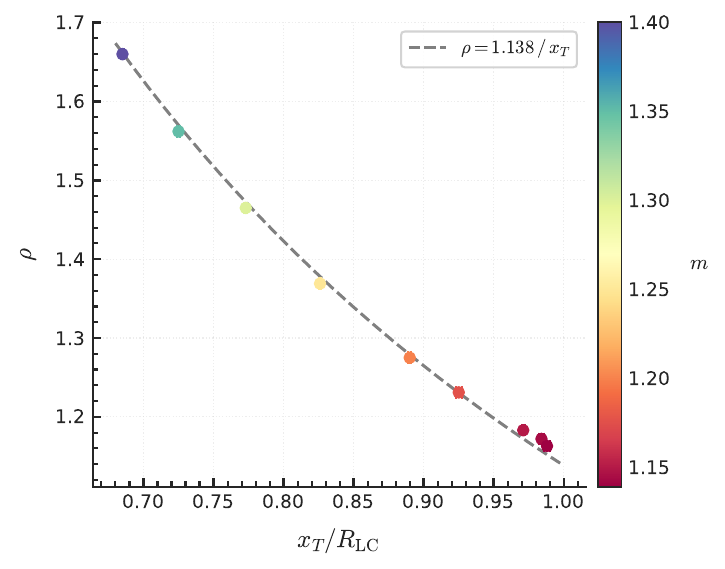}
	\caption{Magnetic flux ratio $\rho$ as a function of the equatorial T-point position $x_T / R_{\text{LC}}$ at polar cap opening angle $\theta_\text{pc} = m\sqrt{R_*}$, with varying $m$. The dashed line represents the curve $\rho = 1.138/x_T$ derived from the baseline configuration for $R_* = 0.25$. Scatter points indicate the average values computed across three independent seeds, with error bars corresponding to standard error.}
	\label{fig9}
\end{figure}

In addition to this physically significant result, we also confirm that the product $\rho \cdot x_T$ remains practically invariant. To visualize this relationship, we turn our attention to the diagnostic plot in Fig. \ref{fig9}, which displays the flux ratio $\rho$ as a function of the T-point position $x_T / R_{\text{LC}}$, providing a direct parallel to the results of Fig. 5 in \citet{dimitropoulos2}. By performing a least-squares fit of the form $\rho = c/x_T$ on the empirical $(\rho, x_T)$ pairs presented in Table \ref{tab3}, we obtain a best-fit parameter of $c = 1.138$. This establishes the curve

\begin{equation}
	\rho = \frac{1.138}{x_T}. \label{eq33}
\end{equation}

\noindent The scatter points in Fig. \ref{fig9} represent the empirical mean results derived for each multiplier $m$, with their respective standard errors plotted as horizontal error bars. Due to the exceptionally low variance across the independent random seeds, these error bars are barely visible on the scale of the plot. Crucially, all empirical data points lie either precisely on or immediately adjacent to the baseline dashed curve. This not only verifies that $\rho \cdot x_T$ remains constant as the flux ratio varies, but it also reveals the true limiting behavior of the system. By extrapolating this curve to the ideal theoretical limit where the T-point touches the light cylinder, we find that the required open flux ratio converges to $\rho \to 1.138$. This limiting value provides a strict, high-precision correction to the standard literature values, e.g., $\rho \approx 1.23$ in \citet{timokhin}.

\section{Summary \& Future Outlook} \label{sec5}

In this paper, we have presented an enhanced PINN-based methodology for solving the pulsar magnetosphere problem, built upon the domain-decomposition framework of \citet{dimitropoulos1}. Our approach replaces the original MLP architecture with RGA KANs, introduces an automated adaptive training pipeline comprising LRA and RBA and embeds physical conditions through hard-constraint ansätze. A modified separatrix update rule and a physics-based convergence criterion eliminate the need for manual supervision and fixed cycle counts, while also yielding more physically accurate solutions.

Focusing on the axisymmetric case allowed us to directly address the specific limitations identified in the baseline framework. A preliminary comparative study based on NTK spectra and geometric complexity showed that the RGA KAN architecture, together with LRA and RBA, yields a significantly better-conditioned optimisation landscape and dramatically compresses the intra-cycle convergence timescale. The combined framework delivers self-consistent magnetosphere solutions with an MSE of the PDE residuals of $\mathcal{O}\left(10^{-6}\right)$ in double precision (two orders of magnitude below the baseline) while reducing total convergence times from several hours to roughly 40 minutes (FP64) or 18 minutes (FP32) on a single GPU. The alignment condition, a known bottleneck in prior work, is minimised to $\mathcal{O}\left(10^{-6}\right)$ within the first training cycle. Our results recover the T-point geometry at the equator; for the baseline polar cap opening angle $\theta_\text{pc} = 1.176\sqrt{R_*}$ at a radius of $R_* = 0.25$, the strict enforcement of pressure balance via the physics-based convergence criterion places the T-point at $x_T = 0.925 \pm 0.002\,R_{\text{LC}}$ in the highest-accuracy runs, providing a significant correction to the earlier reported value of $0.88\,R_{\text{LC}}$.

Ablation experiments confirm that both the advanced architecture and the adaptive weighting strategies are essential for robust, automatic convergence: standard MLPs fail to meet the convergence thresholds, and removing LRA leads to predominantly divergent runs, while deactivating RBA degrades accuracy in the closed-line region by an order of magnitude. Hard-constraint ansätze are likewise indispensable for obtaining physically meaningful magnetospheres. Finally, by systematically decreasing the stellar radius from $R_* = 0.25$ to $0.05$ (an 80\% reduction), we demonstrated that the method can resolve disparate spatial scales. Under a constant flux ratio $\rho$, the T-point remains stationary to within $0.009\,R_{\text{LC}}$. This demonstrates that the poloidal field geometry remains consistently dipolar up to at least $0.25 \ge R_* \ge 0.05$, indicating that the physical size of the star does not influence the global magnetospheric structure within this region. We note, however, that for the smallest star ($R_* = 0.05$), the closed-line PDE residual rises to $\mathcal{O}\left(10^{-3}\right)$; this defines the practical limit of the present implementation.

Based on this result, we fixed the stellar radius at the computationally stable $R_* = 0.25$ (where the framework achieved its highest numerical accuracy) and systematically decreased the polar cap opening angle without enforcing a constant flux ratio. The objective was to push the separatrix outward until the T-point approached the light cylinder ($x_T \to R_{\text{LC}}$). The precision of our methodology allowed us to reach a configuration of $x_T = 0.988 \pm 0.002 \, R_{\text{LC}}$ for a polar cap angle of $\theta_{\text{pc}} = 0.57$ rad and an open flux ratio of $\rho = 1.139$. By extrapolating to the exact limit where $x_T \to R_\text{LC}$, we establish a true asymptotic open flux ratio of $\rho \to 1.138$ (see Eq. \eqref{eq33}). Using this value, the pulsar spindown rate \citep{contopoulos07} evaluates to

\begin{align}
	\dot{E} &= \rho^2 \int_{\Psi/\Psi_S=0}^{\Psi/\Psi_S=1}\frac{3}{2}\left(I\left(\Psi\right)/\, \Psi_S\right)\ \mathrm{d}\left(\Psi/\, \Psi_S\right)\times \dot{E}_{\mathrm{vac}}(90^\circ)\nonumber\\
	&= 1.30\times 0.936 \times \dot{E}_{\mathrm{vac}}(90^\circ) \label{Edot1}\\
	&= 1.217 \dot{E}_{\mathrm{vac}}(90^\circ), \label{Edot}
\end{align}

\noindent where

\begin{equation}
	\dot{E}_{\mathrm{vac}}(\lambda)\equiv B_*^2 R_*^6 \Omega^4\sin^2(\lambda)/(6c^3)
\end{equation}

\noindent denotes the spindown rate of a vacuum dipole rotator with magnetic inclination $\lambda$ and stellar surface magnetic field $B_*$. Notice that the first numerical factor in the r.h.s. of Eq. \eqref{Edot1} comes from the corrected $\rho^2$ factor (1.3 instead of 1.5 established by \citet{spitkovtskyffe}), while the second numerical factor comes from the ratio of the integral of the poloidal electric current distribution obtained in Fig. 5 over the integral of the canonical monopole electric current distribution. This provides a significant correction to the currently widely accepted standard of $\dot{E} \approx 1.5 \dot{E}_{\rm vac}(90^\circ)$ and comes closer to the standard accepted by radio astronomers in evaluating the polar value of the stellar surface magnetic field $B_*$, namely  $\dot{E} \approx \dot{E}_{\rm vac}(90^\circ)$.

The present axisymmetric solution directly resolves the accuracy and efficiency bottlenecks of the baseline method, and the underlying methodology is not restricted to two dimensions. The domain-decomposition strategy, the adaptive training algorithms and the architectural choices carry over to the general three-dimensional, inclined rotator. Naturally, as shown in \citet{dimitropoulos3}, there are certain modifications required, such as replacing the one-dimensional separatrix curve with a two-dimensional surface, or utilizing denser collocation point distributions. Reproducing the results of \citet{dimitropoulos3} with higher accuracy in terms of the pressure balance condition and extending them to arbitrary inclinations are the next natural steps for our proposed framework.

Several additional avenues are opened by this study. For very small stellar radii ($R_* < 0.05$), hierarchical domain decomposition or curriculum training could push the method to even more realistic neutron star scales. Furthermore, while the present study assumes a fixed polar cap geometry, future extensions should aim to self-consistently calculate the exact size and shape of the polar cap for a maximally extending closed-line region. Finally, beyond force-free electrodynamics, the framework could be adapted to relativistic MHD or used for other compact-object magnetospheres with minor modifications.

\section*{Reproducibility statement}

The complete methodology presented in this work is implemented in the open-source GitHub repository titled \texttt{PulsarX}, available at \url{https://github.com/srigas/PulsarX}. The repository includes extensive documentation, installation instructions and usage guidelines. \texttt{PulsarX} is released to ensure full reproducibility of our findings and to provide the scientific community with an accessible, extensible baseline for further refinement of PINN-based magnetosphere solvers and their application to pulsars and other compact-object systems.

\section*{Acknowledgments}

The authors, from the OSCARS-funded project MEXANE, acknowledge the OSCARS project, which has received funding from the European Commission's Horizon Europe Research and Innovation programme under grant agreement No. 101129751. S. R. would like to thank Vasiliki Kousana for proof-reading the original version of the manuscript before submission.

\section*{Generative AI usage statement}

Large Language Models (LLMs) were used throughout this work for grammar, syntax and phrasing refinement only; all ideas, technical content and conclusions remain the authors' work.

\printcredits

\bibliographystyle{cas-model2-names}

\bibliography{cas-refs}


\appendix

\section{RGA KAN architecture} \label{appA}

In a generic KAN layer, the $j$-th output neuron is given by

\begin{equation}
	u_j^{(l)}(\mathbf{x}) = \sum_{i=1}^{d_{l-1}} \sum_{m=1}^{D} w_{jim}^{(l)} \, B_m\!\left(u_i^{(l-1)}(\mathbf{x})\right) + b_j^{(l)},
	\label{eqA1}
\end{equation}

\noindent where $d_{l-1}$ is the input dimension of the layer, $B_m$ a set of basis functions for $m = 1, \dots, D$, and $w_{jim}^{(l)}, b_j^{(l)}$ trainable weights and biases. To define a residual-gated adaptive KAN architecture, two types of basis functions are needed:

\begin{itemize}
	\item \textbf{Chebyshev basis:} $B_m(x) \equiv T_m(\tanh(x))$, where $T_m$ are Chebyshev polynomials of the first kind.
	\item \textbf{Sine basis:} $B^s_m(x) \equiv \sigma^{-1}\left[\sin\left(\omega_m x + p_m\right) - \mu\right]$, where
\end{itemize}

\begin{equation}
	\mu = \exp\left(-\frac{\omega_m^2}{2}\right)\sin\left(p_m\right) \quad \text{and} \quad \sigma = \sqrt{\frac{1}{2} - \frac{1}{2}\exp\left(-2\omega_m^2\right)\cos\left(2p_m\right) - \mu^2} \label{eqA2}
\end{equation}

\noindent are the mean and standard deviation of $\sin\left(\omega_m x + p_m\right)$, respectively.

Within an RGA KAN, the initial $d_I$-dimensional input, $\mathbf{x}$, first passes through a sine-based KAN layer, whose output, $\mathbf{s}\in\mathbb{R}^{d_H}$, is given by

\begin{equation}
	s_j = \sum_{i=1}^{d_I} \sum_{m=1}^{D_s} w_{jim}^{s} \, B^s_m\!\left(x_i\right) + b_j^{s},
	\label{eqA3}
\end{equation}

\noindent where $D_s$ is the number of sine basis functions used. Using this output, two Chebyshev-based KAN layers are defined via:

\begin{equation}
	U_j = \sum_{i=1}^{d_H} \sum_{m=1}^{D} w_{jim}^{u} \, B_m\!\left(s_i\right) + b_j^{u}, \qquad V_j = \sum_{i=1}^{d_H} \sum_{m=1}^{D} w_{jim}^{v} \, B_m\!\left(s_i\right) + b_j^{v},
	\label{eqA4}
\end{equation}

\noindent with $\mathbf{U},\mathbf{V} \in \mathbb{R}^{d_H}$. The outputs of the two gates and the sine-based KAN layer enter the first RGA block. If we assume a total of $N$ identical RGA blocks and denote the input to the $l$-th block by $\mathbf{x}^{(l)}$, with $l = 1, \dots, N$ and $\mathbf{x}^{(1)} \equiv \mathbf{s}$, then the forward pass through each block is defined recursively as:

\begin{align}
	f_j^{(l)} &= \sum_{i=1}^{d_\text{H}}\sum_{m=1}^{D} w^{(l)}_{jim}\, B_m\left(x_i^{(l)}\right) + b^{(l)}_j, \label{eqA5} \\
	g_j^{(l)} &= f_j^{(l)}\, U_j + \left(1 - f_j^{(l)}\right) V_j, \label{eqA6} \\
	z_j^{(l)} &= \beta\, g_j^{(l)} + \left(1-\beta\right)x_j^{(l)}, \label{eqA7} \\
	\tilde{f}_j^{(l)} &= \sum_{i=1}^{d_\text{H}}\sum_{m=1}^{D} \tilde{w}^{(l)}_{jim}\, B_m\left(z_i^{(l)}\right) + \tilde{b}^{(l)}_j, \label{eqA8} \\
	\tilde{g}_j^{(l)} &= \tilde{f}_j^{(l)}\, U_j + \left(1 - \tilde{f}_j^{(l)}\right) V_j, \label{eqA9} \\
	x_j^{(l+1)} &= \alpha\, \tilde{g}_j^{(l)} + \left(1-\alpha\right)x_j^{(l)}, \label{eqA10}
\end{align}

\noindent where the final and intermediate outputs are all $d_H$-dimensional. The final RGA block's output is then fed to the entire architecture's output layer:

\begin{equation}
	y_j = \sum_{i=1}^{d_\text{H}}\sum_{m=1}^{D} w^{o}_{jim}\, B_m\left(x_i^{(N+1)}\right), \label{eqA11}
\end{equation}

\noindent with $\mathbf{y}\in \mathbb{R}^{d_O}$ and $d_O$ corresponding to the output dimension. Based on Eqs. \eqref{eqA2}--\eqref{eqA11}, the total number of trainable parameters of an RGA KAN is:

\begin{equation}
	\left|\boldsymbol{\theta}\right| = 2d_\text{H}\left(d_\text{H}D + 1\right)\left(N+1\right) + 2N + 2D_s
	+ d_\text{H}\left(d_\text{I}D_s + d_\text{O}D + 1\right). \label{eqA12}
\end{equation}

For the purposes of this work, we utilize RGA KANs with 2-dimensional inputs (the spherical coordinates $r,\theta$) for the two regional PINNs: one for the closed-line region with 1-dimensional output, used to approximate $\Psi_\text{cl}$ via Eq. \eqref{eq20}, and one for the open-line region with 2-dimensional output, used to approximate $\Psi_\text{op}$ and $I$ via Eqs. \eqref{eq21}--\eqref{eq22}. We consider a single block ($N=1$) for both RGA KANs, and therefore set $\alpha = \beta = 1$, as the architectures are not very deep. All underlying KAN layers are initialized using the Glorot-like scheme defined in \citet{kaninit}. Finally, following \citet{rga}, we consider $D = 5$ basis functions for all Chebyshev-based KAN layers and $D_s = 8$ basis functions for the sine-based KAN layer at the architecture's input.

\section{Main benchmark in transformed coordinates} \label{appB}

Although the main body of this work treats the pulsar magnetosphere directly in spherical coordinates $(r,\theta)$, we have also solved the baseline benchmark ($R_*=0.25$) in the compactified coordinates used by \citet{dimitropoulos1}, namely

\begin{equation*}
	q = 1/r, \quad \mu = \cos\theta .
\end{equation*}

\noindent This allows us to verify that our accuracy improvements stem from the methodological refinements and not merely from the choice of coordinate representation.

In $(q,\mu)$ coordinates, Eqs. \eqref{eq11}--\eqref{eq13} are transformed as follows. The pulsar equation becomes

\begin{equation}
	\left(q^{2}-\left(1-\mu^{2}\right)\right)\left[ q^{2}\,\frac{\partial^{2}\Psi}{\partial q^{2}} + 2q\,\frac{\partial\Psi}{\partial q} + (1-\mu^{2})\,\frac{\partial^{2}\Psi}{\partial\mu^{2}} \right]
	+ 2\left(1-\mu^{2}\right)\left( q\frac{\partial\Psi}{\partial q} + \mu\frac{\partial\Psi}{\partial\mu} \right) + II'\left(\Psi\right) = 0,
	\label{eqB1}
\end{equation}

\noindent the regularity condition at the light cylinder ($r\sin\theta=1$, which implies $q^{2}=1-\mu^{2}$) reads

\begin{equation}
	II' = -2q^{2}\left( q\frac{\partial\Psi}{\partial q} + \mu\frac{\partial\Psi}{\partial\mu} \right), \quad\text{for}\quad q^{2}=1-\mu^{2},
	\label{eqB2}
\end{equation}

\noindent and the alignment condition takes the form

\begin{equation}
	\frac{\partial\Psi}{\partial q}\frac{\partial I}{\partial\mu} - \frac{\partial\Psi}{\partial\mu}\frac{\partial I}{\partial q} = 0 .
	\label{eqB3}
\end{equation}

To solve Eqs. \eqref{eqB1}--\eqref{eqB3}, we use exactly the same adaptive framework, architectures and double-precision setup described in Sections \ref{sec3} and \ref{sec4}. The ansätze of Eqs. \eqref{eq20} -- \eqref{eq22} are written as:

\begin{align}
	\Psi_{\text{cl}}(q, \mu) &\approx \left(1-\mu^2\right) \, \mathcal{N}_{\text{cl}}(q, \mu), \label{eqB4} \\
	\Psi_{\text{op}}(q, \mu) &\approx \left(1-\mu\right)\Psi_S + \left(1-\mu^2\right) \, \mathcal{N}_{\text{op}, 1}(q, \mu), \label{eqB5} \\
	I_{\text{op}}(q, \mu) &\approx \left(1-\mu^2\right) \, \mathcal{N}_{\text{op}, 2}(q, \mu). \label{eqB6}
\end{align}

\begin{figure}[h!]
	\centering
	\includegraphics[width=0.9\textwidth]{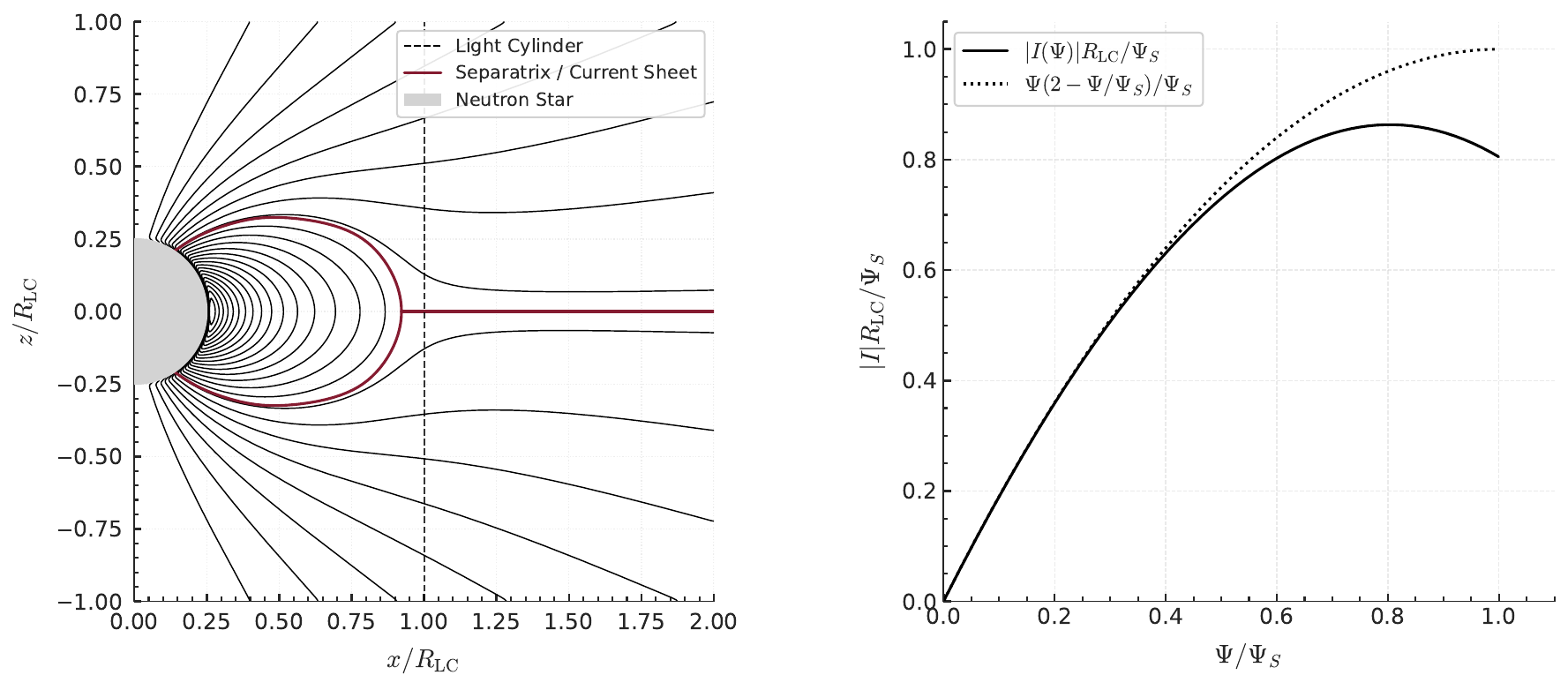}
	\caption{Self-consistent pulsar magnetosphere solution under double precision (FP64), where $z = r\cos\theta$ and $x = r\sin\theta$. Left: Magnetic field lines shown as contours of constant flux. The equatorial T-point is located at $x_T = 0.922 \, R_{\text{LC}}$. Right: Normalized poloidal current distribution $|I(\Psi)| R_{\text{LC}} / \Psi_S$ across the open magnetic field lines. The dotted line denotes the theoretical analytic expression for the split monopole.}
	\label{figB1}
\end{figure}

\noindent Among three runs under independent random seeds, all converge after an average of 15 cycles, with the equatorial T-point located at $x_{T} = 0.922 \pm 0.001\;R_{\mathrm{LC}}$, in agreement with the corresponding spherical-coordinate result to within third-decimal accuracy. The final PDE residual MSEs are $(1.52 \pm 0.66) \cdot 10^{-5}$ for the closed-line region and $(1.75 \pm 0.83) \cdot 10^{-5}$ for the open-line region, which are slightly higher than the $\mathcal{O}\left(10^{-6}\right)$ errors obtained in spherical coordinates but remain below the single-precision spherical results. The magnetic field lines and converged poloidal current as a function of the flux are displayed in Fig. \ref{figB1}; in both cases, the qualitative structure is identical to that obtained in spherical coordinates.

\end{document}